\documentclass[a4paper,twocolumn,fleqn,usenatbib]{mnras}

\usepackage{newtxtext,newtxmath}

\usepackage[T1]{fontenc}
\usepackage{ae,aecompl}

\usepackage{widetext}
\usepackage{graphicx}	
\usepackage{amsmath}	
\usepackage{amssymb}	
\usepackage{epsfig}

\title[Alignment and precession time-scales of a black hole]{Alignment and precession of a black hole misaligned with its accretion disc: Application to Low Mass X-ray Binaries}

\author[S. Banerjee et al.]
{Srimanta Banerjee$^{1}$\thanks{E-mail: srimanta.banerjee@tifr.res.in}, 
Chandrachur Chakraborty$^{2}$ 
and  Sudip Bhattacharyya$^{1}$\\
$^{1}$Department of Astronomy and Astrophysics,
Tata Institute of Fundamental Research, Mumbai 400005, India\\
$^{2}$Kavli Institute for Astronomy and Astrophysics, Peking University, Beijing 100871, China}

\date{Accepted 2019 May 27. Received 2019 May 9; in original form 2019 January 25}

\pubyear{2019}

\begin{document}
\label{firstpage}
\pagerange{\pageref{firstpage}--\pageref{lastpage}}
\maketitle

\begin{abstract}
A thin viscous accretion disc around a Kerr black hole, which is warped due to the Lense-Thirring (LT) effect, was shown to cause the spin axis of the black hole to precess and align with the outer disc. We calculate the total LT torque acting on the black hole, and compute the alignment and precession time-scales for both persistent and transient accretors. In our analysis, we consider the contribution of the inner disc, as it can stay misaligned with the black hole spin for a reasonable range of parameter values.
We find that the alignment time-scale increases with a decrease in the Kerr parameter below a critical Kerr parameter value, contrary to earlier predictions.
Besides, the time-scales are generally longer for transience than the time-scales calculated for persistent accretion. From our analysis of the transient case, we find that the black hole in the low mass X-ray binary (LMXB) 4U 1543--47 could be misaligned, whereas that in the LMXB XTE J1550--564 has aligned itself with the outer disc. The age of the LMXB H 1743--322 is estimated assuming a misaligned disc. We also find that the black hole in a typical Galactic LMXB can take a significantly longer time to align than what was estimated in the past. This may have an important implication on the measurement of black hole spin using the continuum X-ray spectral fitting method.
\end{abstract}




\section{Introduction}\label{intro}
Astrophysical black holes are one of the best laboratories available for probing the observational imprints of general relativity. Lense-Thirring (LT) effect \citep{LT18}, or dragging of local inertial frames due to the rotation of background space-time, is one of such general relativistic effects by which the inner part of a tilted thin accretion disc around a spinning black hole is forced to align with the spin direction of the black hole \citep[Bardeen-Petterson effect;][]{bp}. But the outer part of the disc remains in its original plane as the LT precession rate falls off with the radial distance roughly as $1/R^3$ ($R$ is the radial distance from the centre). The local angular momentum vector of the disc gradually orients itself from the black hole spin direction at the inner part to the misaligned angular momentum direction at the outer part. Hence, the disc becomes warped in between these two regions. However, the misaligned outer portion of the disc does not remain misaligned for ever. It was pointed out by \cite{Rees}, in the context of directional stability of jets in radio-loud active galactic nuclei (AGN), that the torque, which aligns the inner accretion disc with the spin of the black hole, would also align the black hole with the outer part by Newton's third law. He made an estimation of such an alignment time-scale by assuming that the accreting matter carries with it the angular momentum corresponding to the same at the Bardeen-Petterson radius (the radius up to which the disc is aligned), and found the time-scale to be on the order of $10^8$ yr for extremal black holes \citep{Rees,Natarajan}. He thus suggested that the high directional stability of jets in some of the radio-loud AGN arises due to the flywheel effect of the central spinning black hole \citep{Rees,Natarajan}. 
  
The understanding of the physics of warped/tilted accretion disc around a Kerr black hole has considerably improved in the last few decades. The analysis of \cite{bp} along with \cite{Petterson1}, \cite{Petterson2} and \cite{Hatchett} were shown to be inaccurate as they fell short in conserving the angular momentum, and considered the same Shakura-Sunyaev viscosity $\nu=\alpha c_s H$ ($\alpha$ is the Shakura-Sunyaev parameter, $c_s$ is the local sound speed and $H$ is the disc thickness) \citep{Shakura} in all directions \citep{Nixon}. \cite{Pap83} first obtained the accurate evolution equations of a slightly tilted thin accretion disc in the viscous regime, by extending the usual analysis of a flat accretion disc with a new torque component acting on the plane of the disc, as well as from linear hydrodynamical treatment. Later, \cite{Pringle92} obtained the general evolution equation of a warped disc around a central massive object improving the analysis of \cite{Pap83}, and discussed the evolution of surface density as well as radial profile of the disc tilt angle from numerical simulations (with and without LT precession). The analysis of \cite{Pringle92} was also shown to capture the essential physics of a warped viscous disc (although it was not derived from the hydrodynamical analysis of the disc), when the $\alpha$ parameter of the disc, and the tilt angle are small \citep{Ogilvie1999}. 

\cite{SF96} (hereafter, SF96) revisited the problem of alignment of black holes with the outer accretion disc in AGN with the more accurate analysis of \cite{Pringle92}. They assumed both the viscosities and surface density to be constant in the disc, added the LT torque term to the steady state conservation equation of angular momentum density derived in \cite{Pringle92}, and obtained the steady state radial profile of the disc tilt angle, which they used to calculate the alignment time-scale of the black hole. They also showed that the black hole precesses about the angular momentum direction of the outer disc during the process of alignment, and the precession time-scale is identical to the alignment time-scale. However, the contribution of the inner disc was ignored in this work, and they considered the disc to be aligned up to a radius $R_{\rm w}$ (warp radius). They found the time-scale to lie in the range $10^6-3\times10^8$ yr depending upon the warp radius, mass of the black hole and the accretion rate for an extremal black hole. Following \cite{SF96}, \cite{Natarajan} revisited the case of alignment of supermassive black holes in AGN, which was earlier considered by \cite{Rees} (mentioned in a previous paragraph), and they used realistic AGN disc models for this purpose. They found the time-scale of alignment is much less than the derived age of jets in the radio-loud AGN, and thus suggested that the jet directions for such systems are not controlled by the black hole spin, as was earlier proposed to explain the high directional stability of jets found in some of those systems. \cite{Maccarone} discussed the alignment process of black hole spin in microquasars using the analysis of SF96. He showed that the alignment time-scales are to be at least a substantial fraction of the lifetimes of the microquasars GRO J1655--40 and SAX J1819.3--2525. He thus concluded that the inferred misalignment between the disc and the jet in those microquasars was caused by the BP effect. \cite{Martin} improved the analysis of SF96 by considering both the viscosity and surface density to be of radial power law form with the same power law factor (but opposite sign), obtained analytical steady state radial profile of the disc tilt angle, and derived the analytical expressions for alignment as well as precession time-scales. They found these two time-scales to be different contrary to what had been shown by \cite{SF96}, and the precession time-scale to be more sensitive than the alignment time-scale to the change in power law factors. Following the analysis of \cite{Martin}, \cite{Martin1} calculated the alignment time-scale of the micro-quasar GRO J1655--40, and found the lifetime of the mass-transfer rate to be at most a few times the alignment time-scale. Hence, they concluded that the black hole did not get sufficient time to align itself with the outer disc as also found by \cite{Maccarone} with constant viscosity and surface density profiles. \cite{Martin2} used the same analysis of \cite{Martin} to calculate the alignment radius of the accretion disc in the X-ray binary V4641 Sgr (SAX J1819.3--2525) in order to explore the discrepancy between the measurement of jet direction, and inclination of the inner disc from the profile of the Fe K$\alpha$ line. They found, the region of origin of jet and emission site of iron $K\alpha$ line is aligned with the black hole spin, in contradiction with the observation. However, in all the calculations discussed above, the contribution of the inner disc, and the possible change in black hole mass/spin during the alignment process were ignored. Later, \cite{Perego} considered the co-evolution of black hole mass and spin during the alignment process in AGN, and found for a co-rotating disc the fractional increase in mass is typically less than a few percent although the spin modulus can significantly increase up to a few tens of percent. All the discussions so far assume an infinite extension of the accretion disc (or the angular momentum of the disc is much larger than that of a black hole) for studying the alignment process, and consider the black hole-accretion disc system as an open system in which the black hole finally aligns itself with the outer disc's fixed angular momentum. But, if one considers a disc of finite length, and treats this system as a closed system as considered in \cite{King2,Lodato}, the black hole spin direction can be shown to align itself with the direction of total angular momentum of the entire system.

In this paper, we revisit the case of alignment and precession of a black hole in an X-ray binary, and compute the corresponding time-scales analytically and numerically improving upon the previous studies in several ways. The inner part of a warped accretion disc in the viscous regime was always a priori assumed to be aligned with the black hole spin. Thus the contribution of inner disc angular momentum was mostly ignored so far during the process of alignment. But it was shown in \cite{BCB19} (hereafter, BCB19) that for reasonable range of parameter values, the inner disc can remain significantly misaligned with the black hole spin direction. Similar finding was reported earlier in the context of a warped viscous disc in \cite{Lodato}. Recently, X-ray spectral and timing features of the Galactic accreting black hole H1743--322 \citep{Ingram} has also indicated that the inner disc could be tilted. Motivated by this, we consider the contribution of the inner disc, explore how much it affects the time-scales' calculation, and in what parameter regime the deviation from the earlier computed results becomes significant. Most of the black hole X-ray binaries in our Galaxy are transient in nature, and a detailed investigation of the evolution of black hole spin in such a system was not pursued earlier. We study the alignment of black hole in transient systems numerically in this paper, and consider both the surface density and viscosity to be changing during the outburst phase. We also discuss the observational implications of our results in the context of spin measurement of black holes using the continuum X-ray spectral fitting method. In the continuum X-ray spectral fitting method, one fits the thermal continuum spectrum emanating from a black hole accretion disc to the thin disc model of Novikov and Thorne \citep{Novikov}, and find the inner edge of the disc i.e. $R_{\rm ISCO}$. But a reliable measurement of the spin using this method also requires the knowledge of the distance to the black hole, and the inner disc inclination apart from the mass of the black hole. For determining the inner disc inclination, one usually uses the independently measured binary inclination angle \citep{Steiner}. But if the black hole is misaligned with the outer disc angular momentum, measurement of black hole spin using this method can have systematics \citep{Steiner,Narayan}. Hence, it is important to check, before using the continuum X-ray spectral fitting method, whether the black hole has got sufficient time to align itself with the outer disc. Note that the inclination angle is an input parameter for the broad Fe K$\alpha$ line spin measurement method as well \citep{Reynolds}. However, it is in the continuum X-ray spectral fitting method, usually an independently measured inclination angle value is used. Finally, both methods consider an equatorial planar disc, and hence some systematics can be introduced for a tilted inner disc.

In this work, we calculate the alignment time $t_{\rm a}$, i.e., the time the system takes in reducing the misalignment between the black hole spin and outer disc angular momentum direction to $<1\%$ of the initial outer edge tilt angle, for the black holes in low mass X-ray binaries (LMXBs) 4U 1543--475, H1743--322 and XTE J1550--564. These black holes were earlier predicted to be misaligned with the outer disc by \cite{Morningstar}, \cite{Ingram} and \cite{Fragile} respectively. However, \cite{Steiner} proposed that the black hole in XTE J1550--564 has most likely aligned itself with the orbital plane, and placed a constraint on spin-orbital misalignment. We finally calculate the alignment time of a generic Galactic transient black hole LMXB using the averaged description of outbursts reported by \cite{Yu} to appreciate the roughly likelihood of a black hole in our Galaxy to be misaligned, and thus the significance of the assumption of alignment of black hole spin with the outer disc in the context of black hole spin measurement.

The paper is organised as follows. In section \ref{for}, we derive all the equations required for the analyses following roughly the discussions made in BCB19 and SF96, and also develop the numerical scheme used for computing the time-scales. The results obtained from analytical as well as numerical computations are given in section \ref{result}, and a discussions of their implications in LMXBs are investigated in section \ref{lmxb}. We summarize our work in section \ref{con}.
\section{Formalism}\label{for}
First, we discuss the basic physics of a steady state warped accretion disc around a spinning black hole for building the premise on which the subsequent analyses will be done. Then we develop the mathematical framework for studying the precession and alignment processes of the black hole, and discuss the method for obtaining the analytical expressions for the time-scales. Thereafter, we present the scheme that we have considered for obtaining the time-scales numerically for persistent as well as transient accretion. We assume here that the alignment process is much slower than the time the disc takes to attain the steady state (SF96), and the disc is continuously fed with matter carrying angular momentum at the outer edge throughout the process of alignment such that the outer disc angular momentum vector remains fixed (SF96). We will roughly follow the discussion of SF96 and BCB19 for recapitulating the basics of a tilted accretion disc, and building the formalism required to calculate the expressions of time-scales analytically.

\subsection{Warped Accretion disc in Steady State}\label{for1}
We consider a slowly spinning Kerr black hole, with mass $M$ and Kerr parameter $a$, 
surrounded by a geometrically thin (i.e. the disc aspect ratio $H/R<<1$, $H$ is the disk thickness and $R$ is the radial distance from the centre), and Keplerian accretion disc. The spin of the black hole is directed along the $z$ axis in our coordinate system in which the black hole is at the centre. The accretion disc is also initially misaligned with respect to the spin direction of the black hole.

The disc has sufficiently high viscosity, i.e. $\alpha>H/R$, such that the warp is transported diffusively in the disc \citep{Pap83,Ogilvie1999}. We assume the tilt vector $\textbf{l}=(l_x,l_y,l_z)$, the unit vector in the direction of local angular momentum density (i.e. angular momentum per unit area of the disc), makes a small angle with the black hole spin direction such that all the terms beyond first order (e.g., $\vert \partial \textbf{l}/ \partial R \vert^2$) can be neglected in our analysis (SF96). In such a scenario, the angular momentum density conservation equation on each annulus of the disc subjected to the LT torque and viscous torques, takes the following form in the steady state: (SF96, BCB19)
\begin{eqnarray}\label{M}
\frac{1}{R}\frac{\partial}{\partial R}\left[\left(\frac{3R}{L}\frac{\partial}{\partial R}\left(\nu_1 L\right)-\frac{3}{2}\nu_1 \right)\textbf{L}+\frac{1}{2}\nu_2 R L \frac{\partial \textbf{l}}{\partial R}\right]+\frac{\boldsymbol{\omega}_p\times\textbf{L}}{R^3}=0,
\end{eqnarray}
where $\textbf{L}=\sqrt{GMR}\Sigma\textbf{l}$ is the angular momentum density vector, $\nu_1$ and $\nu_2$ are the kinematic viscosities in the disc, and $\omega_p/R^3$ is the LT precession frequency or nodal precession frequency with $\boldsymbol{\omega}_p=(0,0,2GJ_{\rm BH}/c^2)$ ($G$ is the Newtonian gravitational constant, $J_{\rm BH}$ is the angular momentum of the black hole and $c$ is the speed of light in free space) and $\Sigma$ is the surface density of the disc. The viscosity $\nu_1$ corresponds to the azimuthal shear in the disc (i.e., the viscous torque perpendicular to the disc), and it is the usual viscosity associated with accretion discs, responsible for conducting the mass accretion process as well as transporting the $z$ component of angular momentum density. On the other hand, the viscosity $\nu_2$ is associated with the vertical shear (or the viscous torque in the plane of the disc), which appears only if the disc is non-planar, and carries out the diffusion of warps in the disc. We assume the usual Shakura-Sunyaev $\alpha$ parametrization for the viscosity $\nu_1$, and the viscosity $\nu_2$ can then be shown to be related to the $\alpha$ parameter and the viscosity $\nu_1$ of the disc in the following way: \citep{Ogilvie1999}
\begin{eqnarray}\label{alpha}
\frac{\nu_2}{\nu_1}=\frac{1}{\alpha^2}.\frac{2(1+7\alpha^2)}{4+\alpha^2}=f(\alpha).
\end{eqnarray} 
In our formalism, we assume these two viscosities, $\nu_1$ and $\nu_2$, to be independent of the radial co-ordinate, but dependent upon the accretion rate in the following way: \citep{Frank}
\begin{eqnarray}\label{vis}
\nu_1&=&\nu_{10}\dot{M}^{0.3}_{-8},\nonumber\\
\nu_2&=&f(\alpha)\nu_{10}\dot{M}^{0.3}_{-8}=\nu_{20}\dot{M}^{0.3}_{-8},
\end{eqnarray}
where, $\nu_{20}=f(\alpha)\nu_{10}$. The `$-8$' in the subscript implies that the accretion rate is scaled by $10^{-8}\ M_{\odot}\ \rm yr^{-1}$, and $0.3$ in the exponent comes from the Shakura-Sunyaev viscosity parametrization \citep{Shakura}. 

The expression of the steady state angular momentum density can be obtained by taking a scalar product of $\textbf{l}$ with the equation (\ref{M}). Thus, one obtains: (SF96, BCB19) 
\begin{eqnarray}\label{Ld}
L(R)=C_2 R^{1/2}-2 C_1,
\end{eqnarray}
where $C_1$ and $C_2$ are the integration constants. The expressions of $C_1$ and $C_2$ can be determined using the boundary conditions, $\Sigma\rightarrow\Sigma_{\infty}$ as $R\rightarrow\infty$ and $\Sigma(R_{\rm in})=\Sigma_{\rm in}$ ($R_{\rm in}$ is the radius of innermost stable circular orbit, $R_{\rm ISCO}$, for a Kerr black hole) respectively. 
One ultimately arrives by doing the above: \citep[BCB19;][]{CB17}
\begin{eqnarray} \label{C2}
C_2=\sqrt{GM}\Sigma_{\infty},
\end{eqnarray}
and 
\begin{eqnarray}\label{C1} 
C_1=\frac{1}{2}\sqrt{GMR_{\rm in}}\left(\Sigma_{\infty}-\Sigma_{\rm in}\right).
\end{eqnarray}
Here, we note that $C_1$ also depends on the spin of the black hole through $R_{\rm in}$, which is a function of mass and spin of the Kerr black hole \citep{Bardeen}.
If the expression (\ref{Ld}) is used to substitute $L$ in the steady state angular momentum conservation equation (\ref{M}), one arrives at the warped disc equation around a Kerr black hole (BCB19)
\begin{eqnarray}\label{warp1} 
\frac{\partial}{\partial R}\left(3\nu_1 C_1 l_x+\frac{1}{2}\nu_2 R L \frac{\partial l_x}{\partial R}\right)=\omega_p \frac{L}{R^2}l_y,
\end{eqnarray} 
and 
\begin{eqnarray}\label{warp2}
\frac{\partial}{\partial R}\left(3\nu_1 C_1 l_y+\frac{1}{2}\nu_2 R L \frac{\partial l_y}{\partial R}\right)=-\omega_p \frac{L}{R^2}l_x,
\end{eqnarray}  
where  $\boldsymbol{\omega}_p\times\textbf{l}=\left(-\omega_p l_y,\omega_p l_x,0\right)$ in our construction. 
We can also combine the above equations to arrive at (SF96, BCB19)
\begin{eqnarray}\label{warp}
\frac{\partial}{\partial R}\left(3\nu_1 C_1 W+\frac{1}{2}\nu_2 R L \frac{\partial W}{\partial R}\right)=-i\omega_p \frac{L}{R^2}W,
\end{eqnarray} 
where $W=l_x+i l_y=\beta e^{i\gamma}$. Here, $\beta=\sqrt{l_x^2+l_y^2}$ and $\gamma=\tan^{-1}\left(l_y/l_x\right)$ represent the tilt angle and twist angle respectively. These two angles are the Euler angles required for defining an annulus in the disc. By solving the above coupled second order differential equation upon using suitable boundary conditions, one obtains the radial profile of the $x$ and $y$ components of the tilt vector in the steady state, and thus can determine the radial distribution of tilt angle in the disc as a function of the parameters of the system.

As the inner disc alignment was always made as an a priori assumption, the terms associated with $C_1$ was mostly ignored in the warped disc equation, and the attention was mainly drawn towards the evolution of the warped part of the disc. The full warped disc equation was solved numerically and analytically in BCB19 with realistic boundary conditions (disc was truncated at the ISCO radius and the viscous torques at the inner boundary were considered), and it was shown that the inner disc can stay significantly misaligned for a reasonable range of parameter values. The alignment of the inner disc was found to depend mainly upon the parameters $a$, $M$ and $\nu_2$ (or $\nu_1$ and $\alpha$, see equation (\ref{alpha})) as the interplay between the LT torque and viscous torque in the plane of the disc decides the radial profile of the disc tilt angle in the inner disc (check BCB19 for details). It was also shown in BCB19 that a non-zero tilt angle at the inner edge can significantly affect the radial profile of the tilt angle as it increases the strength of the LT torque near the inner edge. Therefore, the inner part of a warped accretion disc can exhibit rich behaviour which was not explored before BCB19.

\subsection{Analytical Computation of Alignment and Precession time-scales}
We study the evolution of black hole spin direction, and develop the mathematical framework for capturing this phenomena in this section. Thus, we discuss the method for computing the alignment and precession time-scales of a black hole analytically on basis of the constructed model.
\subsubsection{Formulation of the model}\label{for2}
In order to study the evolution of black hole spin, we first calculate the total LT torque acting on the black hole, and then discuss how the black hole responds to this torque. We calculate the total LT torque acting on the black hole by computing the same on the entire disc. We work in the black hole frame, and also assume that the outer disc angular momentum vector is fixed as it is continuously fed by matter with fixed angular momentum \citep[SF96;][]{Martin, Perego} throughout this process. The total LT torque acting on the black hole is thus given by 
\begin{eqnarray}\label{dj0}
- \frac{d{\bf J_{\rm BH}}}{dt}&=&\int \frac{{\bf \omega_p} \times {\bf L}}{R^3}~2\pi R dR \nonumber\\
&=&\int \frac{2\pi \omega_p L(R)}{R^2}(-l_y(R),l_x(R),0)dR. 
\end{eqnarray}
Here, $\textbf{J}_{\rm BH}=(J_x,J_y,J_z)$ is the total black hole angular momentum. Thus, one obtains, (here, $R_g=GM/c^2$ and $\beta_f$ is the tilt angle at the outer edge) 
\begin{eqnarray}\label{dj1}
\frac{d(J_{x} + i J_{y})}{dt}=4\pi a c R_g^2\beta_f (I_1-i I_2), 
\label{dj}
\end{eqnarray}
where, 
\begin{eqnarray}
I_1&=&\frac{1}{\beta_f}\int \frac{L(R)}{R^2}l_y(R) dR,\label{int1}\\ 
I_2&=&\frac{1}{\beta_f}\int \frac{L(R)}{R^2}l_x(R) dR.  \label{int2}
\end{eqnarray}
Since, the black hole spin orientation is changing in this process, an observer in the black hole frame will see the outer disc tilt angle to change, and finally to become aligned with the black hole spin axis. By a simple coordinate transformation, one can connect the outer edge tilt angle with the black hole spin orientation \citep[SF96;][]{Martin} and, it gives (SF96)
\begin{eqnarray}\label{coordinate}
\beta_f&=&-(j_x+i j_y).
\end{eqnarray}
where $j_x=J_x/J_{\rm BH}$ and $j_y=J_y/J_{\rm BH}$. Using this result, one can arrive at the following expression of evolution of black hole spin axis
\begin{eqnarray}\label{bhang}
j_x+i \ j_y=\beta_f \exp\left[-\frac{4\pi R_g}{M}(I_1-i I_2)t\right].
\end{eqnarray}
Hence, we see the black hole precesses with exponentially decreasing amplitude, and finally aligns itself with the disc outer edge angular momentum. The real part of the above defines the alignment process whereas the imaginary part exhibits the precession of a black hole around the outer disc angular momentum direction. Here, we note that the left hand side (lhs) of the equation (\ref{dj0}), which describes evolution of black hole spin axis, is analysed in the frame of fixed outer disc whereas the right hand side (rhs) of the equation is calculated with respect to an observer in the black hole frame. Thus, one must perform a frame transformation on the rhs in order to obtain the components of the black hole angular momentum vector in the outer disc frame \citep{Perego}. But for small outer edge tilt angle, the rotation matrix of such a transformation can be shown to be nearly identical to the identity matrix \citep{Perego}.

Therefore, the alignment time-scale $T_{\rm align}$ and precession time-scale $T_{\rm precess}$ are given by 
\begin{eqnarray}
T_{\rm align}&=&\frac{M}{4\pi R_g I_1},\label{time1}\\
T_{\rm precess}&=&\frac{M}{4\pi R_g I_2}.\label{time2}
\end{eqnarray}
So we find that the alignment time-scale depends on the $l_y$ profile or $x$ component of torque, whereas the precession time-scale depends on the $l_x$ profile or $y$ component of the LT torque. For computing the integrals $I_1$ and $I_2$, we need the radial profiles of $l_y$ and $l_x$ respectively. We use the analytically calculated radial profiles obtained in BCB19 for this purpose, and discuss the method for computing the expression of these integrals in the next section.
\subsubsection{Determination of the expressions of time-scales}\label{an}
In this section, we present the analytical calculation of the integrals $I_1$ and $I_2$ mentioned in equations (\ref{int1}) and (\ref{int2}). We will follow the formalism discussed in BCB19 for calculating the radial profiles of the $x$ and $y$ components of tilt vectors analytically, and then we derive the expressions of the corresponding components of LT torque by computing the integrals $I_1$ and $I_2$.   

We briefly discuss the formalism which we use for obtaining the radial profiles of $l_x$ and $l_y$ for the sake of completeness before computing the integrals. We use perturbative approach for solving the warped disc equations (\ref{warp1},\ref{warp2}), and expand $l_x$ and $l_y$ in orders of the Kerr parameter $a$ in the following way \citep[BCB19;][]{CB17}:
\begin{eqnarray}\label{exp}
l_x&=&l_x^{(0)}+a l_x^{(1)}+a^2 l_x^{(2)}.....,\nonumber \\
l_y&=&l_y^{(0)}+a l_y^{(1)}+a^2 l_y^{(2)}.....
\end{eqnarray}
Here, in our formalism for spinning black holes, $l_x^{(0)}$ and $l_y^{(0)}$ are the $x$ and $y$ components of the tilt vector when $a\simeq0$, but not for exactly $a=0$ (for practical reasons; see BCB19). We ignore the contribution of $l_y^{(0)}$ as in the limit of $a\rightarrow0$ (or LT torque tending to zero) the twist angle becomes very small giving negligible contribution to $l_y^{(0)}$ (see the expression of twist angle below equation (\ref{warp})). Before solving the warped disc equations of different orders, we first make the equations (\ref{warp1},\ref{warp2}) dimensionless using $R_g$ as the length scale and $C_1$ as the scale for angular momentum density (please check section 3.1.2 of BCB19 for relevant discussions), and arrive at the following the expressions: (BCB19)
\begin{eqnarray}\label{noeq}
R\frac{\partial^2 l_x}{\partial R^2}+\left[(n+1)\frac{C_1}{L}+3/2\right]\frac{\partial l_x}{\partial R}&=&4 a \xi \frac{l_y}{R^2},\\ \nonumber
R\frac{\partial^2 l_y}{\partial R^2}+\left[(n+1)\frac{C_1}{L}+3/2\right]\frac{\partial l_y}{\partial R}&=&-4 a \xi \frac{l_x}{R^2},
\end{eqnarray}
where, $n=\frac{6\nu_1}{\nu_2}$ and $\xi = \frac{cR_g}{\nu_2}$. 

In this work, we consider the expansions of $l_x$ and $l_y$ up to second order for computing the integrals $I_1$ and $I_2$. Hence, we need the expressions of $l_x^{(0)}$ , $l_x^{(1)}$, $l_x^{(2)}$, $l_y^{(1)}$ and $l_y^{(2)}$. Thus, we solve the warped disc equations (\ref{noeq}) up to second order in Kerr parameter to arrive at the expressions of the terms mentioned in the previous paragraph. Excepting $l_y^{(2)}$, all the terms mentioned above were derived in BCB19 (see section 3.1 of BCB19 for the expression of the above mentioned terms). Here, we only state the expression of $l_y^{(2)}$ without going into the details of calculation. For more details, the reader is referred to the section 3.1 of BCB19. The expression of $l_y^{(2)}$ can be obtained by solving the warped disc equations (equation (\ref{noeq})) of second order upon using the boundary conditions that $l_y^{(2)}$ is zero at $R=R_{\rm in}$ and as $R\rightarrow\infty$ (as we do not expect the outer edge of the disc to be affected by frame-dragging effect). This gives  
\begin{eqnarray}\label{ly2}
&&l_y^{(2)}=Q_4\left(\frac{\sqrt{R}}{L}\right)^n+\frac{1}{n \left(n^2-4\right)R}.
\left[Q  \left(n \left(C^2 R+\right.\right.\right.\nonumber \\
&&4 C \sqrt{R} \left(z^n-1\right)
\left.\left.\left.+4-4 z^n\right)-2 C^2 R+2 n^2 \left(C \sqrt{R}-1\right).\right. \right.\nonumber\\
&&\left.\left.\left(z^n+1\right)\right)+Q_3 \left(n^2-4\right) R\right],\nonumber\\
\end{eqnarray}
where,
\begin{eqnarray}
&&Q_4=\frac{-Q (-C)^{n+2}-2 Q_3 (-C)^n-Q_3 n (-C)^n}{n (n+2)}, \ \ \ Q =\frac{4 \xi \beta^{r}_{a,\rm in} }{1-z_{\rm in}^n},\nonumber
\end{eqnarray}
and 
\begin{eqnarray}
&&Q_3=\frac{n}{1-z^n_{\rm in}}\left[\frac{C^2 . Q . z^n_{\rm in}}{n(n+2)}-\frac{Q}{n \left(n^2-4\right)R_{\rm in}}. \right. \nonumber \\
&&\left.\left(n \left(C^2 R_{\rm in}
+4 C\sqrt{R_{\rm in}} \left(z_{\rm in}^n-1\right)-4 z_{\rm in}^n+4\right)-2 C^2 R_{\rm in}\right.\right.\nonumber \\
&&\left.\left.+ 2 n^2 \left(C \sqrt{R_{\rm in}}-1\right) \left(z_{\rm in}^n+1\right)\right)\right],\nonumber \\
\end{eqnarray}
where, $\beta_{a, \rm in}^i$ is the value of $l_x^{(1)}$ at the inner disc edge. Also, 
\begin{eqnarray}\label{par}
&&z=1+\frac{2C_1}{L}=\frac{z_{\rm in}\sqrt{R}}{z_{\rm in}\sqrt{R}+(1-z_{\rm in})\sqrt{R_{\rm in}}},\ \& \ z_{\rm in}=z(R_{\rm in})=\frac{\Sigma_{\infty}}{\Sigma_{\rm in}}. \nonumber \\
\end{eqnarray}
We note that the above expansion (equations (\ref{exp})) is valid up to a critical value of Kerr parameter $a_c$ beyond which the higher order terms become more dominant than the zeroth order term resulting in the breakdown of this formalism. This critical value can be shown to depend on values of the viscosity component $\nu_2$, and the mass of the black holes (for more details see the section 4.2 of BCB19).

We now use the expressions of $l_x$ and $l_y$ of different orders to calculate the integrals $I_1$ and $I_2$. The integrals $I_1$ and $I_2$ are given by:
\begin{eqnarray}\label{i1}
I_1&\simeq&\frac{1}{\beta_f}\int \frac{L(R)}{R^2}\left(al_y^{(1)}(R)+a^2 l_y^{(2)}(R)\right) dR\nonumber\\
&=& a I_1^{a}+a^2 I_1^{a^2}
\end{eqnarray}
and 
\begin{eqnarray}\label{i22}
I_2&\simeq&\frac{1}{\beta_f}\int \frac{L(R)}{R^2}\left(l_x^{(0)}(R)+al_x^{(1)}(R)+a^2l_x^{(2)}(R)\right) dR\nonumber\\
&=& I_2^{0}+a I_2^{a}+a^2 I_2^{a^2}
\end{eqnarray}
where,
\begin{eqnarray}
&&I_1^{a}=\frac{1}{\beta_f}\int \frac{L(R)}{R^2}l_y^{(1)}(R) dR,\
I_1^{a^2}=\frac{1}{\beta_f}\int \frac{L(R)}{R^2}l_y^{(2)}(R) dR,\nonumber \\
&&I_2^{0}=\frac{1}{\beta_f}\int \frac{L(R)}{R^2}l_x^{(0)}(R) dR,\
I_2^{a}=\frac{1}{\beta_f}\int \frac{L(R)}{R^2}l_x^{(1)}(R) dR,\ \nonumber
\end{eqnarray}
and 
\begin{eqnarray}
I_2^{a^2}&=&\frac{1}{\beta_f}\int \frac{L(R)}{R^2}l_x^{(2)}(R) dR.
\end{eqnarray}
The expression of these integrals can be obtained analytically in MATHEMATICA on using analytical expressions of $l_x$ and $l_y$ of different orders.

Thus we derive the expressions of the integrals $I_1$ and $I_2$ considering up to second order expansions in $x$ and $y$ components of tilt vector. One can find the expressions of the time-scales easily from the equations (\ref{time1}) and (\ref{time2}) upon using the integral values of $I_1$ and $I_2$. The expressions for time-scales take the following form as one makes the integrals $I_1$ and $I_2$ dimensionless using the scheme mentioned before:
\begin{eqnarray}\label{nondim}
T_{\rm align}&=&\frac{M}{4\pi C_1 I_1},\nonumber\\
T_{\rm precess}&=&\frac{M}{4\pi C_1 I_2}.
\end{eqnarray}
To appreciate the functional dependence of the time-scales on $\nu_2$ (or $\xi$; see the expression of $\xi$ below equation (\ref{noeq})), $\alpha$, and accretion rate $\dot{M}$, we write $C_1$ in the following way using the equations (\ref{alpha}, \ref{C1}, \ref{par}) 
\begin{eqnarray}\label{C1n}
C_1=\frac{\xi\dot{M}}{12\pi \alpha^2}.\sqrt{R_{\rm in}}.\left(1-\frac{1}{z_{\rm in}}\right).
\end{eqnarray}
Here, we substitute the surface density at infinity $\Sigma_{\infty}$ in equation (\ref{C1}) by $\dot{M}$ and $\nu_1$ by the following relation \citep{Frank,Natarajan}
\begin{eqnarray}\label{sur}
\Sigma_{\infty}=\frac{\dot{M}}{3\pi\nu_1}.
\end{eqnarray}
As also mentioned earlier, our analytical expressions of the time-scales are relevant up to $a_c$, as beyond this value the expansion of tilt vector does not remain to be valid.
\subsection{Numerical Computation of Alignment and Precession time-scales}
In this section, we discuss the basic scheme for computing the alignment and precession time-scales numerically. First, we consider the numerical framework for studying the persistent accretion in which the accretion rate is fixed throughout the alignment process, and then move on to the case of transient accretion where the accretion rate cycles between quiescent and outburst phases during the alignment process. We work in the frame of black hole for studying the evolution of black hole spin axis. Thus, we determine the time-scales by monitoring how the outer edge tilt vectors evolve with time.
\subsubsection{Persistent accretion}\label{per1}
For computing the time-scales for persistent accretion, we first divide the entire evolution time of black hole spin axis into several short time intervals. The viscous time-scale associated with $\nu_1$ can be considered to be the time the disc takes to achieve the steady-state \citep{Frank,Martin}. Our short time-interval is always chosen larger compared to this viscous time-scale.

As the black hole starts precessing and aligning itself with the outer disc angular momentum, an observer in the black hole frame would find components of the outer edge tilt vectors to change, and thus the outer edge boundary conditions would also change resulting in different radial profiles of $l_x(R)$ and $l_y(R)$. Hence, the total torque acting on the disc/black hole will also change accordingly. Therefore, we calculate the total LT torque on the disc (or black hole) at each time-interval, and use this information to find new values of $l_x$ and $l_y$ at the outer edge (i.e., the new direction of black hole spin). Then we use this new values of $l_x$ and $l_y$ at the outer edge as an initial condition for calculating the total torque for the next step. We continue this process until the outer edge tilt angle becomes $e^{-1}$ times the initial tilt angle for computing the alignment time-scale, and we run this process until $\tan^{-1}(l_y/l_x)$ becomes very close to $1$ for computing the precession time-scale. This is a natural way to define the time-scales as it comes from the fact that the evolution of $x$ and $y$ components of the black hole spin (or $l_x$ and $l_y$ at the outer edge w.r.t an observer in the black hole frame) are described by the functions $\beta_f e^{-t/T_{\rm align}}\cos\left(t/T_{\rm precess}\right)$ and $\beta_f e^{-t/T_{\rm align}}\sin\left(t/T_{\rm precess}\right)$ respectively (see the expression (\ref{bhang})). We also calculate the square of the deviation between numerically obtained values of $l_x$ and $l_y$ and that obtained from the theoretical estimations (i.e., we assume the evolution of $l_x$ and $l_y$ in each  short time-interval can be approximated by the above mentioned functions) at each interval, and obtain the time-scales by minimizing the sum of the squares for checking the reliability of our numerical scheme.

In order to calculate the total LT torque acting on the black hole at each time-interval, we need to solve the integrals (\ref{dj0}), and for this purpose we require the radial profile of the $x$ and $y$ components of the tilt vector. Thus we solve the warped disc equations (\ref{noeq}) numerically subjected to the following boundary conditions: (BCB19)
\begin{eqnarray}\label{b1}
l_x(R_{\rm in})&=&\beta_i \cos(\gamma_i),\nonumber\\
l_y(R_{\rm in})&=&\beta_i \sin(\gamma_i),
\end{eqnarray}
and 
\begin{eqnarray}\label{b2}
l_x(R_{f})&=&\beta_f, \nonumber\\
l_y(R_{f})&=&0,
\end{eqnarray}
where, $R_f$, $\beta_i$, $\beta_f$ are the outer edge radius, tilt angles at the inner edge and outer edge of the disc respectively. We ignore the contribution of $l_y$ at the outer edge as the twist angle is very small there (see the expression of twist angle below equation (\ref{warp})). The above boundary conditions are defined for the first iteration, and as the black hole spin evolves the outer boundary conditions would change accordingly. To determine the outer boundary conditions after the first iteration, we transform the lhs of the equation (\ref{dj0}) into the corresponding components of the tilt vector at the outer edge by using the transformation law mentioned in equation (\ref{coordinate}). This gives,
\begin{eqnarray}\label{num}
&&\Delta j_x +i \Delta j_y = -(\Delta l_x(R_f)+i \Delta l_y(R_f))\nonumber \\
&&= \frac{4\pi C_1 }{M}\Delta t \left(\int \frac{L(R)}{R^2}l_y(R) dR- i \int \frac{L(R)}{R^2}l_x(R) dR\right) \nonumber \\
&&=\frac{4\pi C_1 }{M}\Delta t(\mathcal{I}_1-i  \mathcal{I}_2)
\end{eqnarray}
where, $\mathcal{I}_1=\beta_f I_1$ and $\mathcal{I}_2=\beta_f I_2$. Here, the integrals 
$\mathcal{I}_1$ and $\mathcal{I}_2$ are in dimensionless form (please see the paragraph above equation (\ref{noeq}) for the scheme used for making the integrals dimensionless). 
Thus, the above equation (\ref{num}) describes the evolution of $l_x$ and $l_y$ at the outer edge, and one can use this for calculating the time-scales in the way discussed in the previous paragraph.
\subsubsection{Transient accretion}\label{tran}
\begin{figure}
\begin{center}
\includegraphics[width=0.49\textwidth]{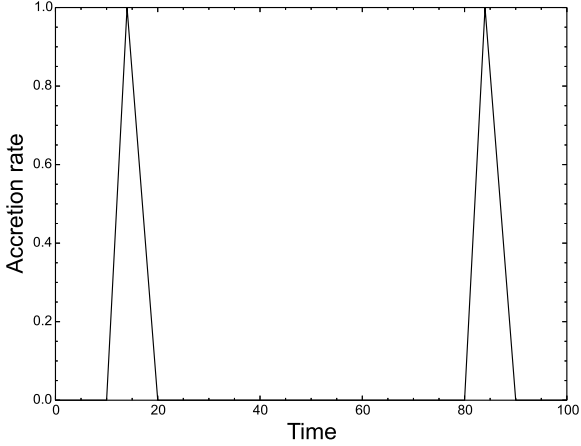}
\caption{\label{fg0} A schematic illustration of accretion rate evolution for the transient accretion. The accretion rate is normalized by the peak accretion rate, and the time is in arbitrary unit. Here, two outbursts are shown with an identical triangular profile separated by the quiescent phase (see section \ref{tran} for more details).\label{fig0}}
\end{center}
\end{figure}
Most of the Galactic black hole binaries are transient in nature \citep{Tetarenko, Belloni}. A transient behaviour of such a system is characterized by episodic accretion onto the black hole. Most of the time, such a black hole stays in the quiescent phase lasting possibly from a few months to several decades, and the accretion rate in such a phase is extremely low (typical X-ray luminosity, $L_X<10^{33}\rm erg \ s^{-1}$ \citep{Belloni}). Such an inactive period is interrupted by episodic outbursts, in which the X-ray flux (or accretion rate) increases by several orders of magnitude, presumably as a consequence of thermo-viscous disc instability \citep{Lasota}. Since, the change in black hole spin direction strongly depends on the accretion rate (see the equation (\ref{num}) and expression of $C_1$ in (\ref{C1n})) and the accretion rate is very low in the quiescent phase compared to the outburst phase, we ignore the change in the orientation of black hole spin axis during the quiescent phase for all practical purposes. Therefore, it is a reasonable to assume a zero accretion rate in the quiescent phase.
Thus, the accretion phenomena for a transient black hole system for the computation of time-scales can be modelled as a process consisting of a series of alternate outbursts and quiescent phases \citep{Maccarone} in which the accretion can be safely assumed to occur only during the outburst phase.

We consider the accretion process in the outburst phase as a linear increase in the accretion rate from zero in the quiescence to a maximum value $\dot{M}_{\rm max}$, and then linear decrease to quiescence as shown in Fig. \ref{fg0} \citep{BC16}. Hence, for an outburst phase with a duty cycle of $f$, we can write, \citep{BC16}
\begin{eqnarray}\label{avg}
\dot{M}_{\rm av}\simeq\frac{1}{2}f\dot{M}_{\rm max}
\end{eqnarray}
Here, the factor $1/2$ comes from our assumption of representing the outburst phase as a triangular profile of accretion rate, and the duty cycle defines the ratio between the outburst time and the recurrence time for a transient source.

For calculating the time-scales for transient accretion, we divide an outburst in small time intervals, and for each time interval we consider a constant accretion rate with the disc being assumed to be in steady state. But, the accretion rate changes from one time interval to another, as the outburst rises or decays.  Thus within a time-interval, our formalism for transient accretion is identical to the persistent case. During the outburst phase, the viscosities in the accretion disc would change with the surface density from one time interval to another as the accretion rate changes. The viscosities in this outburst state are assumed to change according to equation (\ref{vis}) with the accretion rate. The change in the surface density is captured through the change in the viscosity and accretion rate by the relation (\ref{sur}). This assumption of steady disc within a time interval holds if the viscous time-scale (in which matter is accreted through the disc) is less than the accretion rate rise and decay times of the outburst. The change in the  black hole spin direction in each time-interval during the outburst cycles is calculated following the numerical scheme described in the section \ref{per1}. Here,
we also assume that all the outbursts during the process of alignment can be described by same profile of accretion rate for a particular black hole. 

The case of transient accretion was earlier considered in \cite{Maccarone}.
\cite{Maccarone} approximated the accretion history of a transient system as a series of outburst events in which matter accretes with Eddington accretion rate and $30$ percent efficiency (i.e, the accretion rate in the outburst phase was described by a sharp peak with Eddington accretion rate instead of a more realistic triangular profile shown in Fig. \ref{fg0}), and reframed the SF96 result for transient accretion. Therefore, the time-scale for transient case came out just as the time-scale calculated using SF96 result multiplied with a factor coming from the ratio between recurrence and outburst times (i.e., inverse of duty cycle). Besides, \cite{Maccarone} did not consider the accretion rate dependent viscosity, and overestimated the alignment time-scale by a factor of $50$ due to a mistake in the calculation \citep{Steiner,Erratum}.
\section{Theoretical results and Discussions}\label{result}
In this section, we first compare our analytical results with the numerical ones for a persistent source. We discuss then our numerical results for persistent accretion in detail, and
explore the quantitative differences between our results and the same obtained in SF96 as a function of the parameters of the system. In this context, we identify the regime in the parameter space in which the deviation from the results discussed in SF96 is prominent. Thereafter, we present our numerical results for transient accretion, and study in detail how the time-scales change relative to our persistent results for same parameter values.  
\subsection{Parameter Values}\label{param}
In order to explore the behaviour of time-scales as functions of the parameters $\nu_2$, $\alpha$, $M$ and $\dot{M}$, we have to choose suitable ranges of these quantities relevant to the case of black hole X-ray binaries. We choose mass of the black hole in the range $5-15M_{\odot}$ \citep{Fragos}, and the accretion rate for such black holes in the range $1-0.001$ $\dot{M}_{\rm Edd}$ \citep{Yu} where $\dot{M}_{\rm Edd}=L_{\rm Edd}/\epsilon c^2$ is the Eddington accretion rate ($\epsilon=0.1$ is the efficiency of the accretion flow and $L_{\rm Edd}$ is the Eddington luminosity). Besides, we consider a range $0.05-0.4$ \citep{King1} for $\alpha$ (correspondingly the range $0.03-0.94$ for $n$; see the expression of $n$ below equation (\ref{noeq}), and equation (\ref{alpha})) and $10^{14}-2\times10^{15}$ $\rm cm^2\ s^{-1}$ \citep{Frank} for the viscosity component $\nu_{20}$. The range for the viscosity component $\nu_{10}$ can be determined from the range of $\nu_{20}$ and $\alpha$ using equation (\ref{alpha}).  We also set the inner and outer edge tilt angles to $0^{\circ}$ and $10^{\circ}$ respectively, and $z_{\rm in}=\Sigma_{\infty}/\Sigma_{\rm in}$ to $0.5$ for the purpose of demonstration. For this value of inner edge tilt angle, the value of inner edge twist angle is not required for studying the evolution of black hole spin direction (see the boundary conditions (\ref{b1})). Here, we take into account the case of prograde rotation ($a>0$) only.
\subsection{Comparison between analytical and numerical results}\label{anper}
\begin{figure*}
\centering
\includegraphics[width=0.49\textwidth]{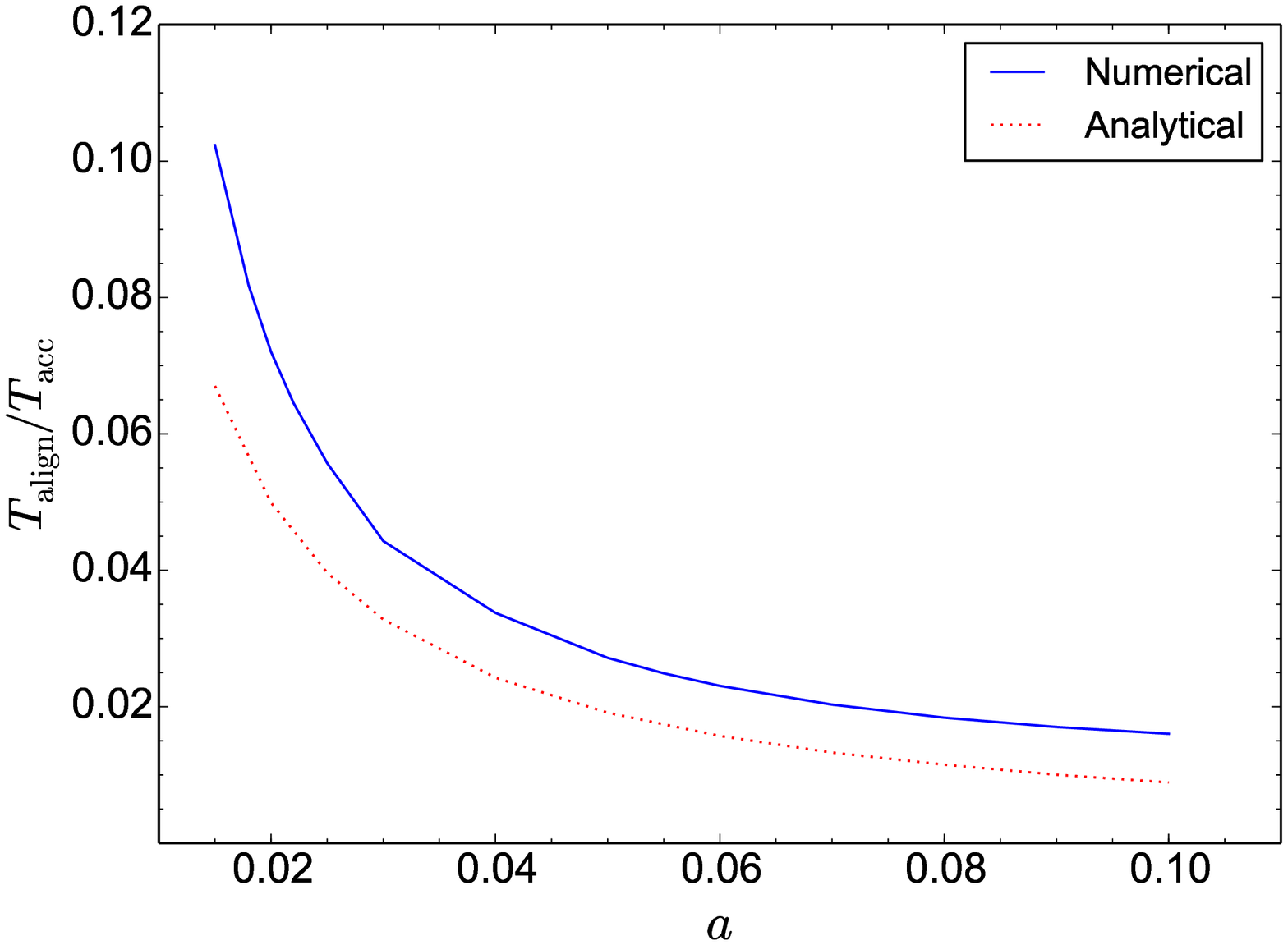}
\includegraphics[width=0.49\textwidth]{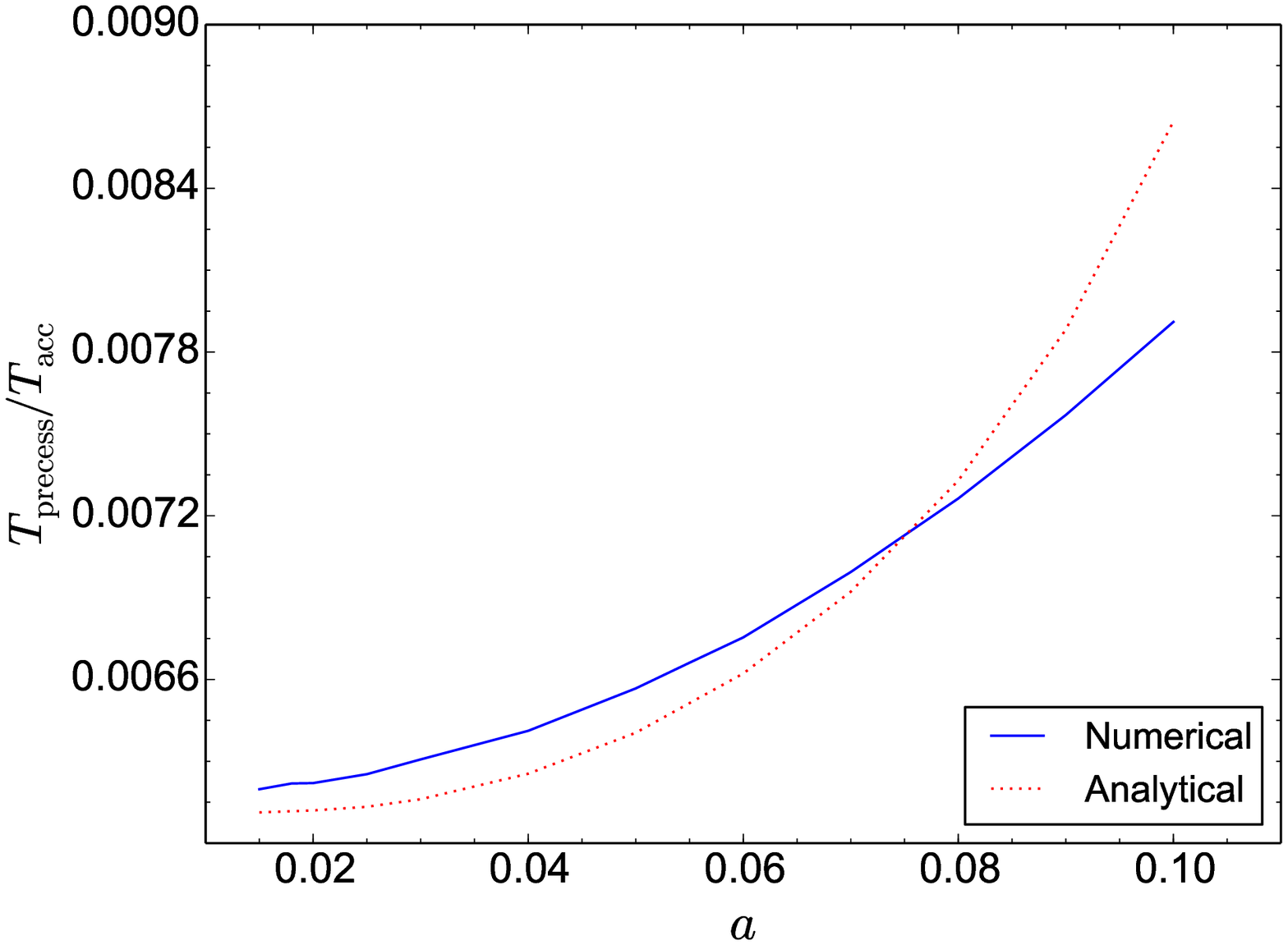}
\caption{\label{fg1}Alignment (Left panel) and Precession (Right panel) time-scales for different values of the Kerr parameter. Other parameter values are: $\dot{M}=10^{-8}\ M_{\odot}\ \rm yr^{-1}=0.1\ \dot{M}_{\rm Edd}$ for $M=5\ M_{\odot}$, $\alpha=0.15$, and $\nu_{20}=10^{15}\ \rm cm^2\ \rm s^{-1}$. Here, $T_{\rm acc}=\frac{M}{\dot{M}}$ is the accretion time-scale. The term `Numerical' signifies our numerically obtained result, and `Analytical' denotes the result obtained by our analytical calculations (see section \ref{anper} for more details).\label{fig1}}
\end{figure*}
We compare our numerical results for the persistent case with the results obtained using our analytical calculations. We find that the behaviour of time-scales as functions of the Kerr parameter match qualitatively (see Fig. \ref{fig1}). Interestingly, the alignment time-scale decreases with the increase in the Kerr parameter, opposite to the behaviour exhibited by the precession time-scale.

In order to appreciate the behaviour of time-scales as functions of the Kerr parameter, one should check the equations (\ref{i1}) and (\ref{i22}). The expression of the integral $I_1$ (see the equation (\ref{i1})) contains two terms, $aI_1^a$ and $a^2I_1^{a^2}$, and both these terms decrease as $a$ decreases. 
Thus, alignment time-scale (equation (\ref{nondim})), being inversely proportional to $I_1$, increases as $a$ takes a lower value (see left panel of Fig \ref{fig1}). But, the integral $I_2$  (see the equation (\ref{i22})) contains three terms, $I_2^0$, $aI_2^a$ and $a^2I_2^{a^2}$, and the term $a^2I_2^{a^2}$ ($\mid I_2^{a^2}\mid>>I_2^{a}$) is negative unlike the other terms. Thus, as $a$ takes a higher value, the term $a^2I_2^{a^2}$ reduces the total contribution of all the terms in $I_2$ more than the increase in $aI_2^a$ resulting in decrease of $I_2$. As a result, precession time-scale (equation (\ref{noeq})) increases as it is inversely proportional to $I_2$ (see right panel of Fig. \ref{fig1}). Here, we note that the quantities $I_1^a$, $I_1^{a^2}$, $I_2^0$, $I_2^a$, $I_2^{a^2}$ and $C_1$ also depend on $a$ through $R_{\rm in}$ (see equation (\ref{C1}); $R_{\rm in}$ is the ISCO radius for a Kerr black hole). But their changes with decreasing/increasing $a$ are relatively small.

As mentioned before, our analytical result is valid up to a critical value of the Kerr parameter (`$a_c$'), and there also exists an upper limit `$a_u$' of the Kerr parameter up to which the analytical results match within $10\%$ with the numerical result (BCB19). This upper limit as well as $a_c$ can be shown to be a function of mass of the black hole and $\nu_2$ (see section 4.2 of BCB19 for more details). For the choice of the parameters considered in Fig. \ref{fg1}, $a_u$ can be shown to be roughly around $0.1$ ($a_c$ is close to $0.13$), which justifies the range of Kerr parameters used for comparison in those panels. 
\begin{figure*}
\centering
\includegraphics[width=0.49\textwidth]{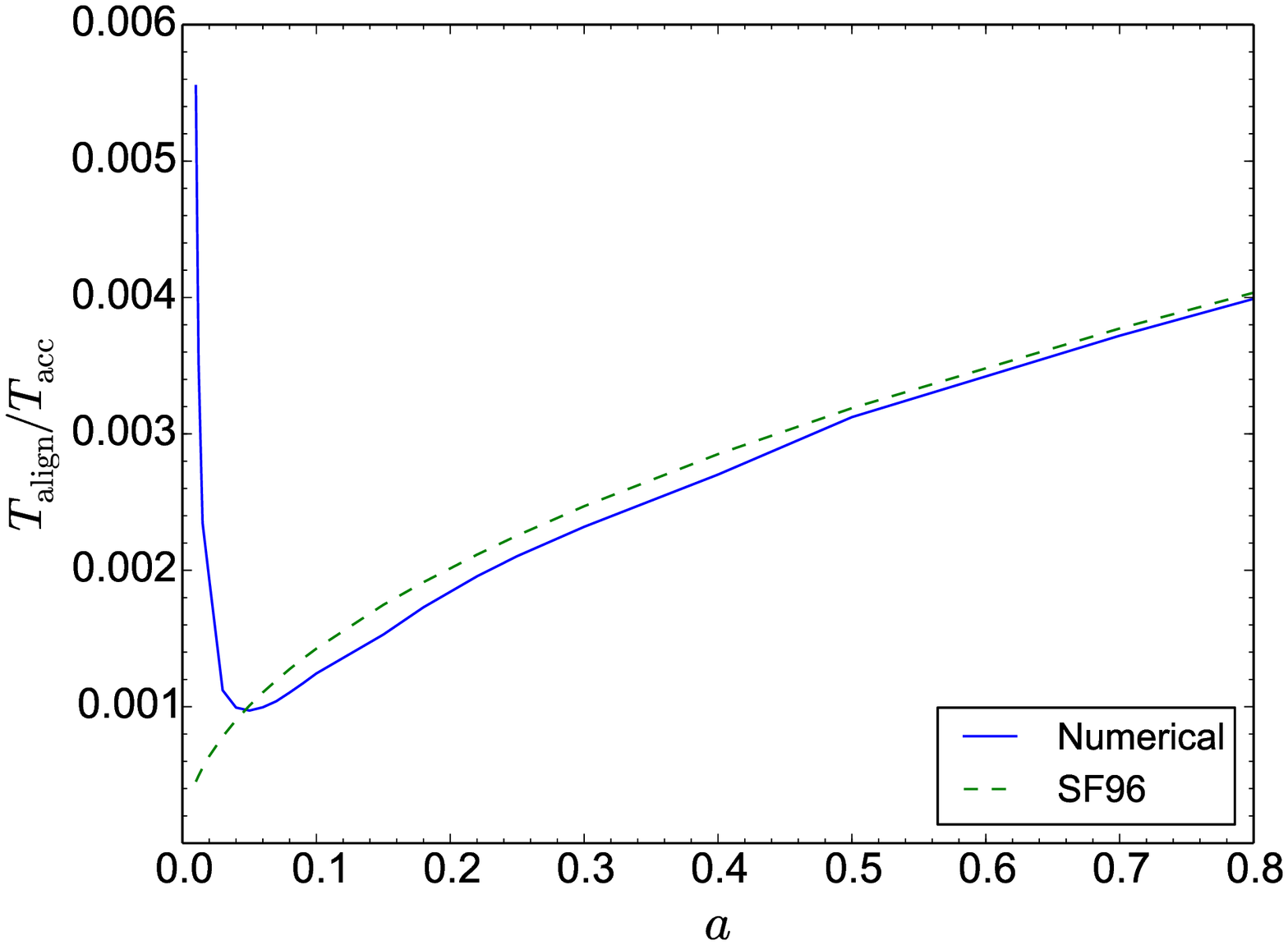}
\includegraphics[width=0.49\textwidth]{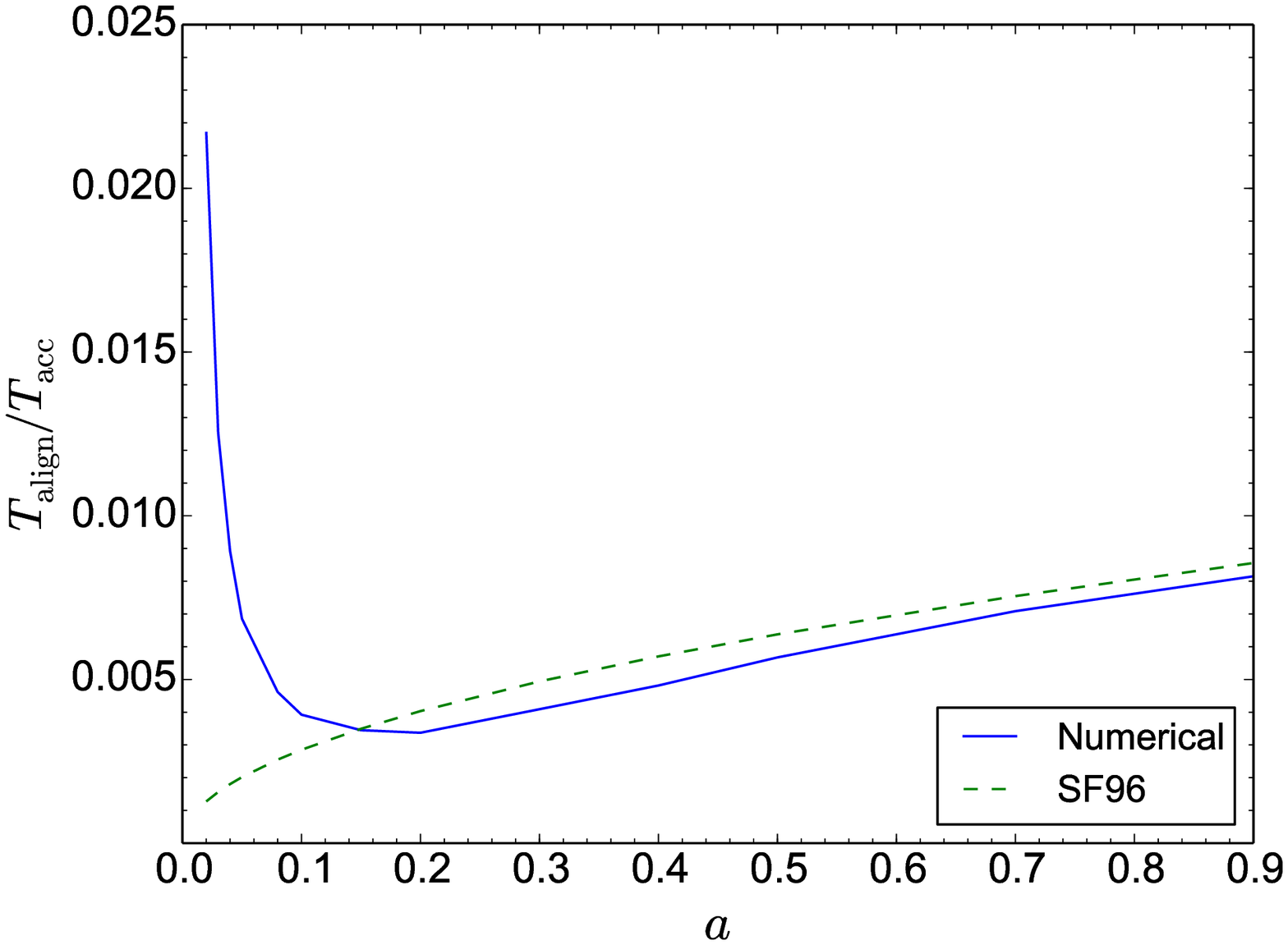}
\caption{\label{fg2}Alignment time-scale for different values of the Kerr parameter for the viscosities $\nu_{20}=5\times10^{14}\ \rm cm^2\ \rm s^{-1}$ (Left panel) and $\nu_{20}=2\times10^{15}\ \rm cm^2\ \rm s^{-1}$ (Right panel). Other parameter values are: $\dot{M}=10^{-8}\ M_{\odot}\ \rm yr^{-1}=0.05\ \dot{M}_{\rm Edd}$ for $M=10\ M_{\odot}$, and $\alpha=0.1$. Here, accretion time-scale, $T_{\rm acc}=\frac{M}{\dot{M}}$. The term `Numerical' signifies our numerically obtained result, and `SF96' denotes the result obtained by SF96 (see section \ref{per} for more details).\label{fig2}}
\end{figure*}
\subsection{Numerical Results: Persistent Accretion}\label{per}
We investigate in detail our numerical results for the persistent accretion as a function of the parameters $\nu_2$, $a$ and $\alpha$ in comparison to the same obtained by SF96. According to the calculations in SF96, the alignment time-scale is identical to the precession time-scale, and the expression of them has the following form: \citep[SF96;][]{Natarajan}
\begin{eqnarray}\label{sf96}
T_{\rm align}=T_{\rm precess}=\frac{3}{\zeta}.\frac{M}{\dot{M}}.\left(\frac{a\xi}{2}\right)^{1/2}
\end{eqnarray}
where, $\zeta=cR_g/\nu_1$. Here, one should note that SF96 did not take into account the contribution of the inner disc, and essentially assumed that the surface density to be independent of radius. 
\begin{figure*}
\centering
\includegraphics[width=0.49\textwidth]{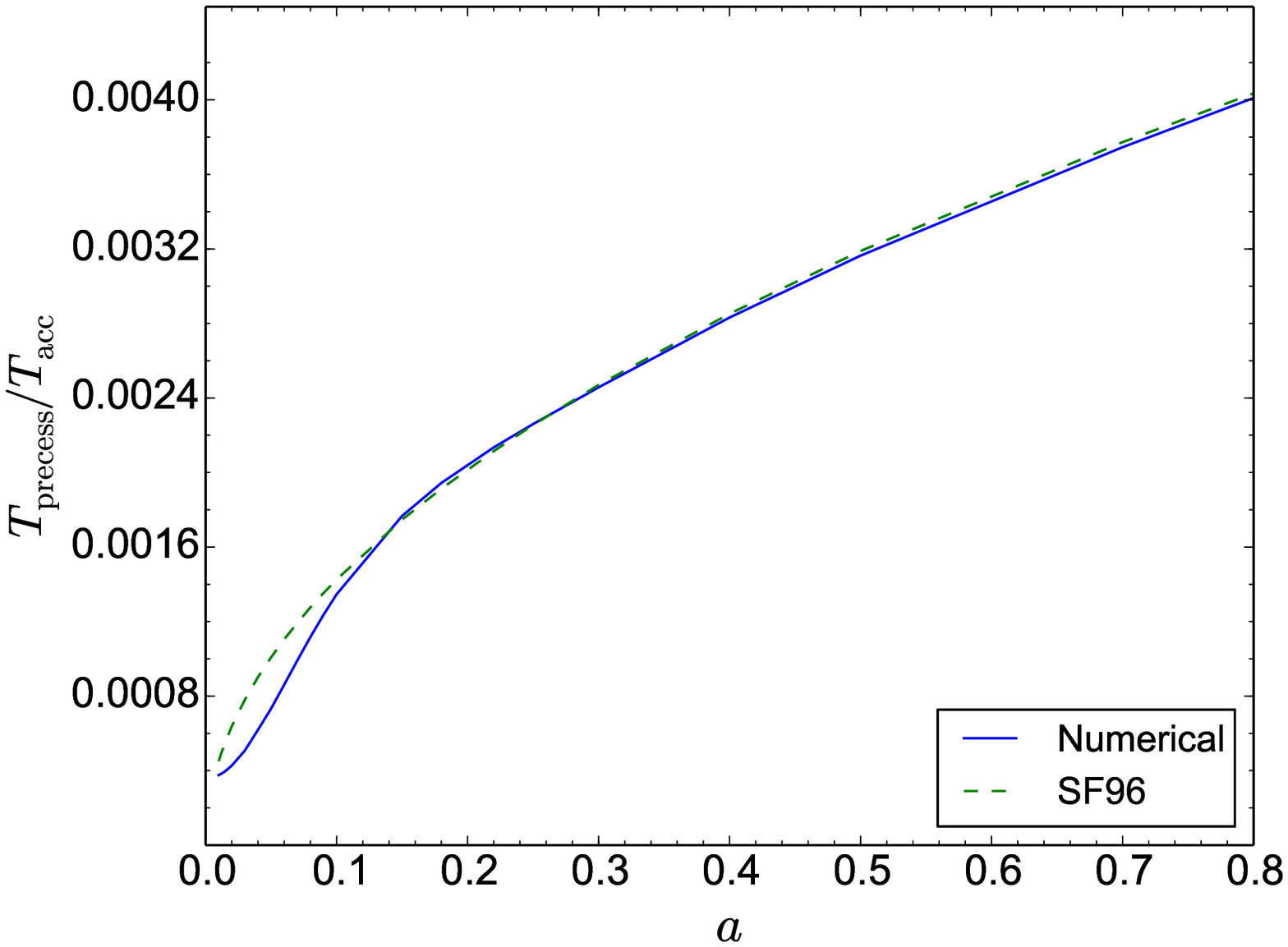}
\includegraphics[width=0.49\textwidth]{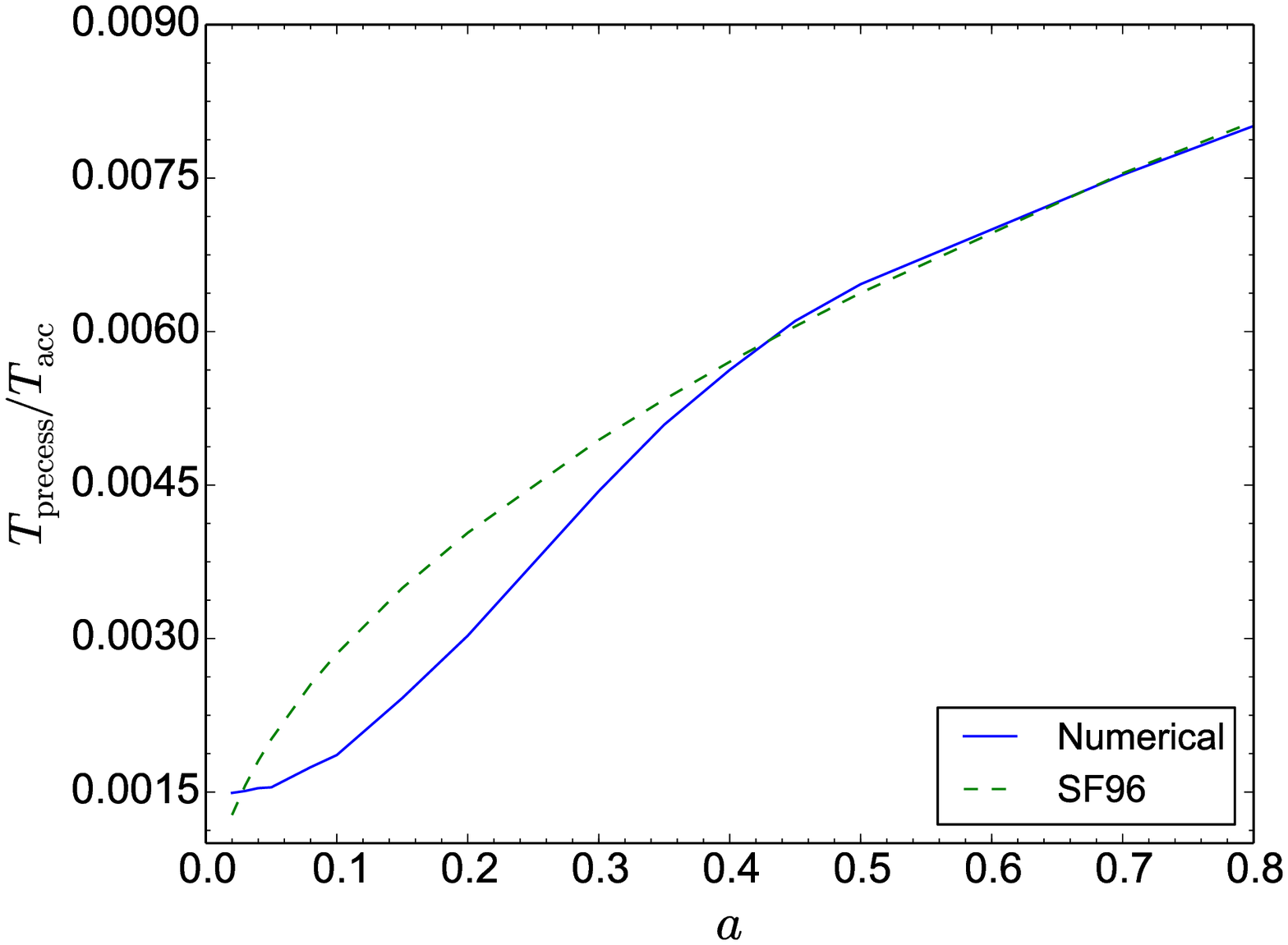}
\caption{\label{fg3}Precession time-scale for different values of the Kerr parameter for the viscosities $\nu_{20}=5\times10^{14}\ \rm cm^2\ \rm s^{-1}$ (Left panel) and $\nu_{20}=2\times10^{15}\ \rm cm^2\ \rm s^{-1}$ (Right panel). Other parameter values are: $\dot{M}=10^{-8}\ M_{\odot}\ \rm yr^{-1}=0.05\ \dot{M}_{\rm Edd}$ for $M=10\ M_{\odot}$, and $\alpha=0.1$. Here, accretion time-scale, $T_{\rm acc}=\frac{M}{\dot{M}}$. The term `Numerical' signifies our numerically obtained result, and `SF96' denotes the result obtained by SF96 (see section \ref{per} for more details).\label{fig3}}
\end{figure*}
\begin{figure*}
\centering
\includegraphics[width=0.49\textwidth]{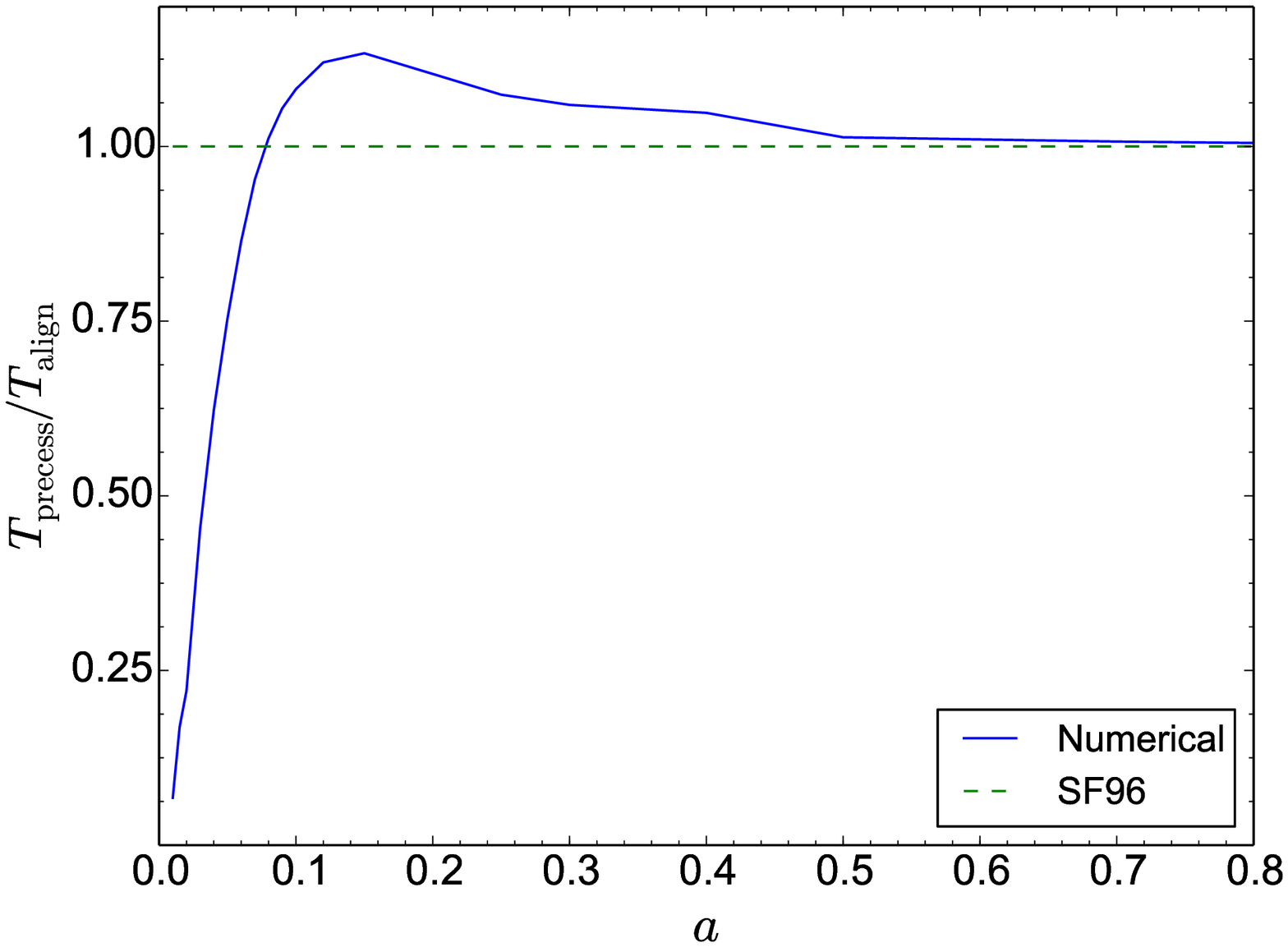}
\includegraphics[width=0.49\textwidth]{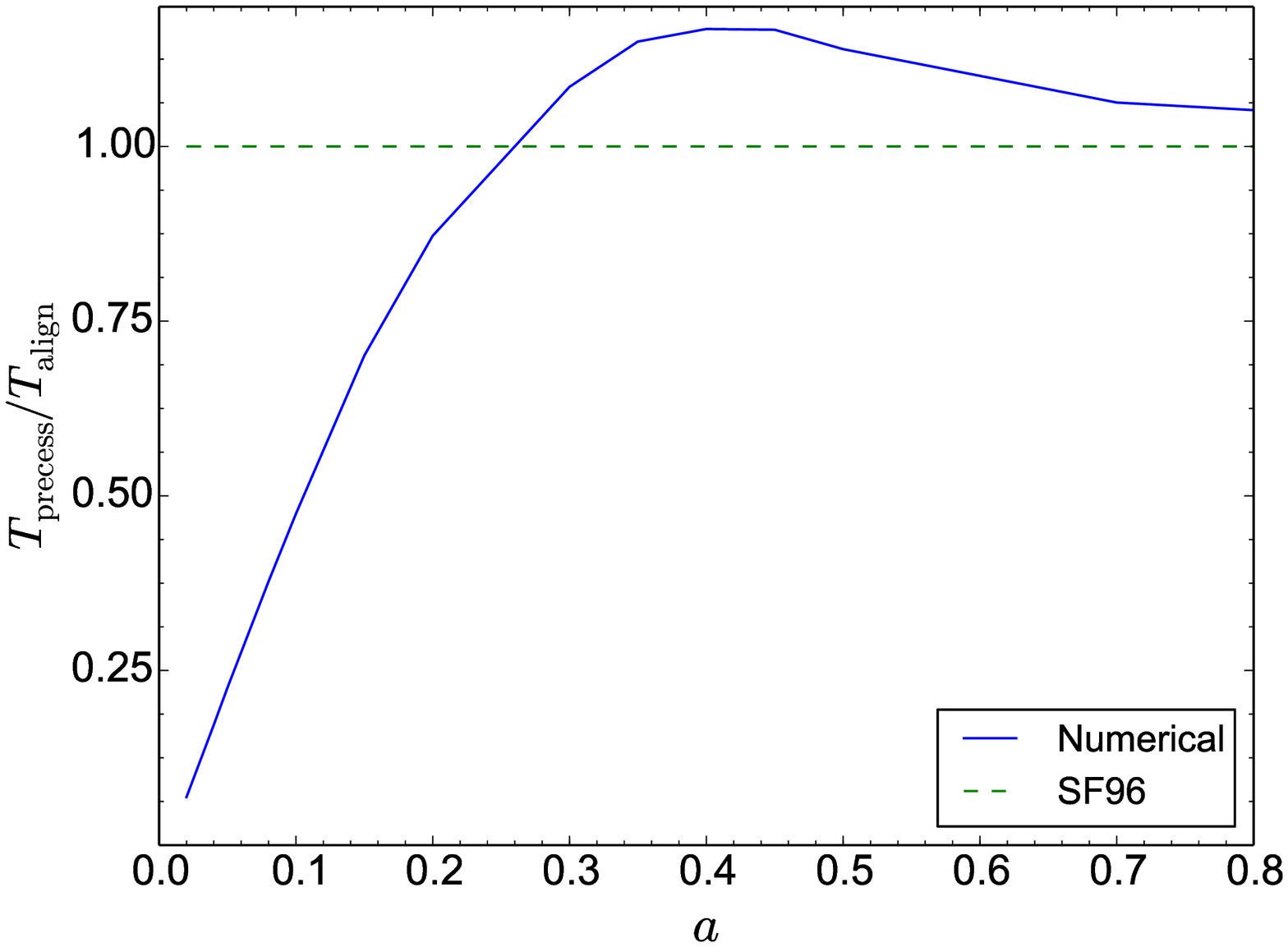}
\caption{\label{fg4}Ratio between precession and alignment time-scales for different values of the Kerr parameter for the viscosities $\nu_{20}=5\times10^{14}\ \rm cm^2\ \rm s^{-1}$ (Left panel) and $\nu_{20}=2\times10^{15}\ \rm cm^2\ \rm s^{-1}$ (Right panel). Other parameter values are: $\dot{M}=10^{-8}M_{\odot} \rm yr^{-1}=0.05\ \dot{M}_{\rm Edd}$ for $M=10\ M_{\odot}$, and $\alpha=0.1$. Here, accretion time-scale, $T_{\rm acc}=\frac{M}{\dot{M}}$. The term `Numerical' signifies our numerically obtained result, and `SF96' denotes the result obtained by SF96 (see section \ref{per} for more details).\label{fig4}}
\end{figure*}
\begin{figure*}
\centering
\includegraphics[width=0.49\textwidth]{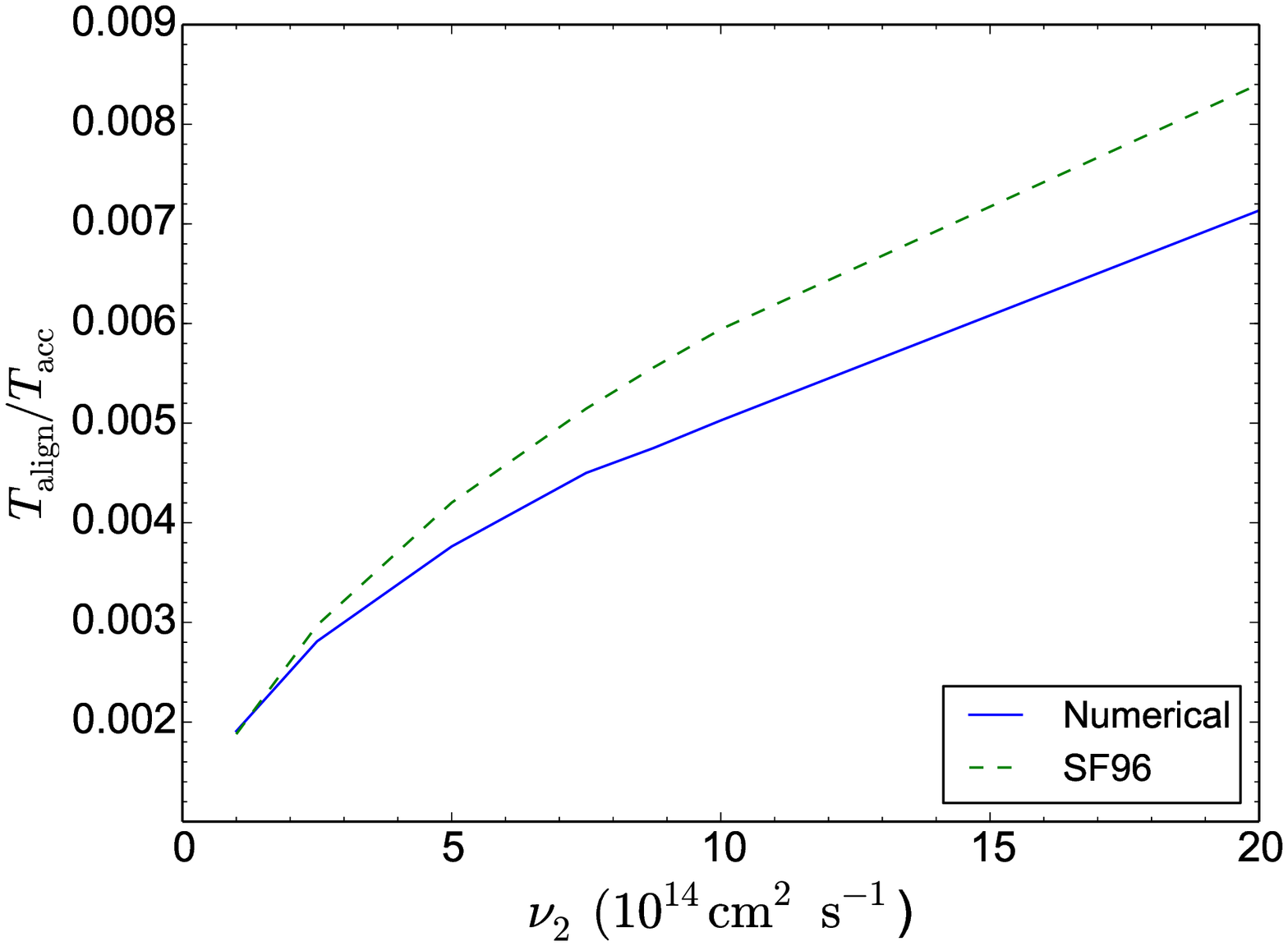}
\includegraphics[width=0.49\textwidth]{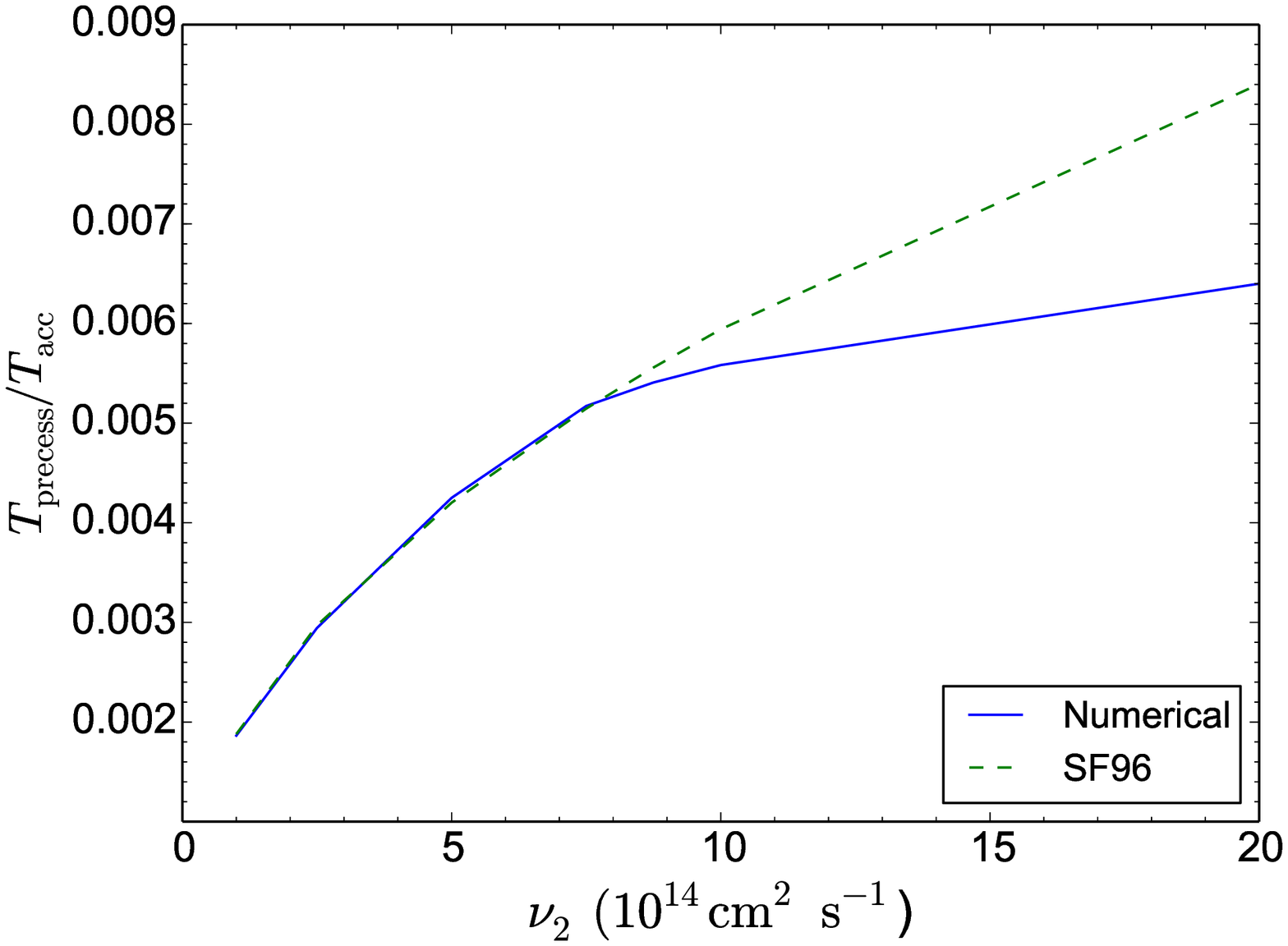}
\caption{\label{fg5}Alignment (Left panel) and Precession time-scales (Right panel) for different values of viscosity component $\nu_2$. Other parameter values are: $\dot{M}=10^{-8}\ M_{\odot} \rm yr^{-1}=0.05\ \dot{M}_{\rm Edd}$ for $M=10\ M_{\odot}$, $a=0.2$ and $\alpha=0.1$. Here, accretion time-scale, $T_{\rm acc}=\frac{M}{\dot{M}}$. The term `Numerical' signifies our numerically obtained result, and `SF96' denotes the result obtained by SF96 (see section \ref{per} for more details).\label{fig5}}
\end{figure*}

The behaviour of alignment time-scale for the low values of Kerr parameter in our computation is qualitatively different from that obtained using the expression (see equation (\ref{sf96})) mentioned in SF96 (Fig. \ref{fg2}). In our case, the alignment time-scale increases with the decrease in the Kerr parameter below a certain critical value (`$a_{T}$') as opposed to what predicted by SF96. For $a>a_{T}$, our results match well with that of SF96, and the time-scale value roughly follows $a^{1/2}$ behaviour (as exhibited by the equation (\ref{sf96})) discussed in SF96 (Fig. \ref{fg2}). This critical value $a_{T}$ depends on the mass of the black hole as well as the viscosity parameter $\nu_2$ (see Fig. \ref{fg2}). For example, when $M=10\ M_{\odot}$ and $\nu_2=5\times10^{14}\ \rm cm^2\ \rm s^{-1}$, $a_{T}$ is around $0.05$, whereas for $\nu_2=2\times10^{15}\ \rm cm^2\ \rm s^{-1}$, $a_{T}\sim 0.22$, and also for $M=5M_{\odot}$ and $\nu_2=10^{15}\ \rm cm^2\ \rm s^{-1}$ (we consider these parameter values in Fig. (\ref{fg1})), we find $a_{T}\sim 0.21$ (this is why the $a^{1/2}$ behaviour does not show up in Fig. \ref{fg1} as $a$ is considered up to $0.1$).

As one can see from the equation (\ref{num}), the alignment time-scale depends on $C_1$ and the integral $\mathcal{I}_1$. The value of $\mathcal{I}_1$ depends mainly upon the radial profile of $l_y$ (equation \ref{int1}), while the quantity $\xi$ (see the definition below equation(s) (\ref{noeq})) essentially controls the behaviour of $l_y$ through the warped disc equations (equation \ref{noeq}). Thus, the behaviour of the time-scale depends strongly upon $\xi$, and the dependence of $a_{T}$ on the black hole mass/viscosity comes through this quantity. Here, we note that although $C_1$ also depends on $\nu_2$ and the mass of black hole (see equation (\ref{C1n})), the quantity $\mathcal{I}_1$ dictates the behaviour of the alignment time-scale as a function of the Kerr parameter.

Precession time-scale in our calculation also deviates from the functional form derived in SF96 (equation (\ref{sf96})) for the low Kerr parameter values, while maintaining the same trend, i.e., increasing with the Kerr parameter (Fig. \ref{fg3}). The value of the Kerr parameter, beyond which the precession time-scale roughly starts following the $a^{1/2}$ form, also depends on $\nu_2$ (see Fig. \ref{fg3}) and mass of the black hole like for the alignment time-scale case. The $\nu_2$ dependence of both the time-scales are studied 
in Fig. \ref{fg5}, and one essentially finds that increasing the value of $\nu_2$ increases the deviation from the SF96 result (as described by the equation \ref{sf96}). We also note that the alignment time-scale can have different values from the precession time-scale in contradiction to what predicted by SF96, and the ratio between these time-scales slowly converges to unity for higher values of the Kerr parameter. The value of Kerr parameter, beyond which both the time-scales are identical, also depends on the mass of the black hole and viscosity because of the reason discussed in the previous paragraph (see Fig. \ref{fg5}).

The above mentioned behaviour of the time-scales can be appreciated in terms of the parameter, warp radius $R_{\rm w}$. It is defined as the distance at which the local LT precession time-scale ($R^3/\omega_p$) becomes equal to the viscous time-scale ($R^2/\nu_2$).
Thus, it is given by (see the expressions of $\xi$ and $\omega_p$ below equations (\ref{noeq}) and (\ref{M}) respectively)  (SF96, BCB19)
\begin{eqnarray}\label{warprad}
R_w=\frac{\omega_p}{\nu_{2}}=2a\xi R_g.
\end{eqnarray}
The warp radius $R_w$ roughly defines the region up to which the effect of LT precession dominates, and it is also directly related to the alignment radius $R_{\rm align}$ which is roughly $0.094\ R_w$ (see BCB19 for details). So, the inner disc may not be aligned with the black hole spin when $R_w$ is close to the black hole inner radius $R_{\rm in}$. Therefore, when $R_w$ is large compared to the disc inner edge radius ($R_{\rm w}>>R_{\rm in}$), the inner disc is aligned with the black hole spin, and our results are expected to match with the predictions of SF96. Our results for the higher values of Kerr parameter (i.e., higher values of $R_w$ when $M$ and $\nu_2$ are fixed as in Figs. \ref{fg1} and \ref{fg2}) or lower $\nu_2$ value (i.e. higher values of $R_w$ when $M$ and $a$ are fixed as in Fig. \ref{fg5}) are in line with the above argument. On the other hand, when $R_{\rm w}$ is close to $R_{\rm in}$, the deviation is prominent as in this limit the contribution of the inner disc can be significant \citep[BCB19;][]{Lodato}, and our results expectedly start departing from the prediction of SF96 (as it was ignored in SF96). Besides, as this warp radius depends on the Kerr parameter, mass of the black hole and the viscosity $\nu_2$, it plays an important role in controlling the behaviour of time-scales.
\begin{figure*}
\centering
\label{fig6}
\includegraphics[width=0.49\textwidth]{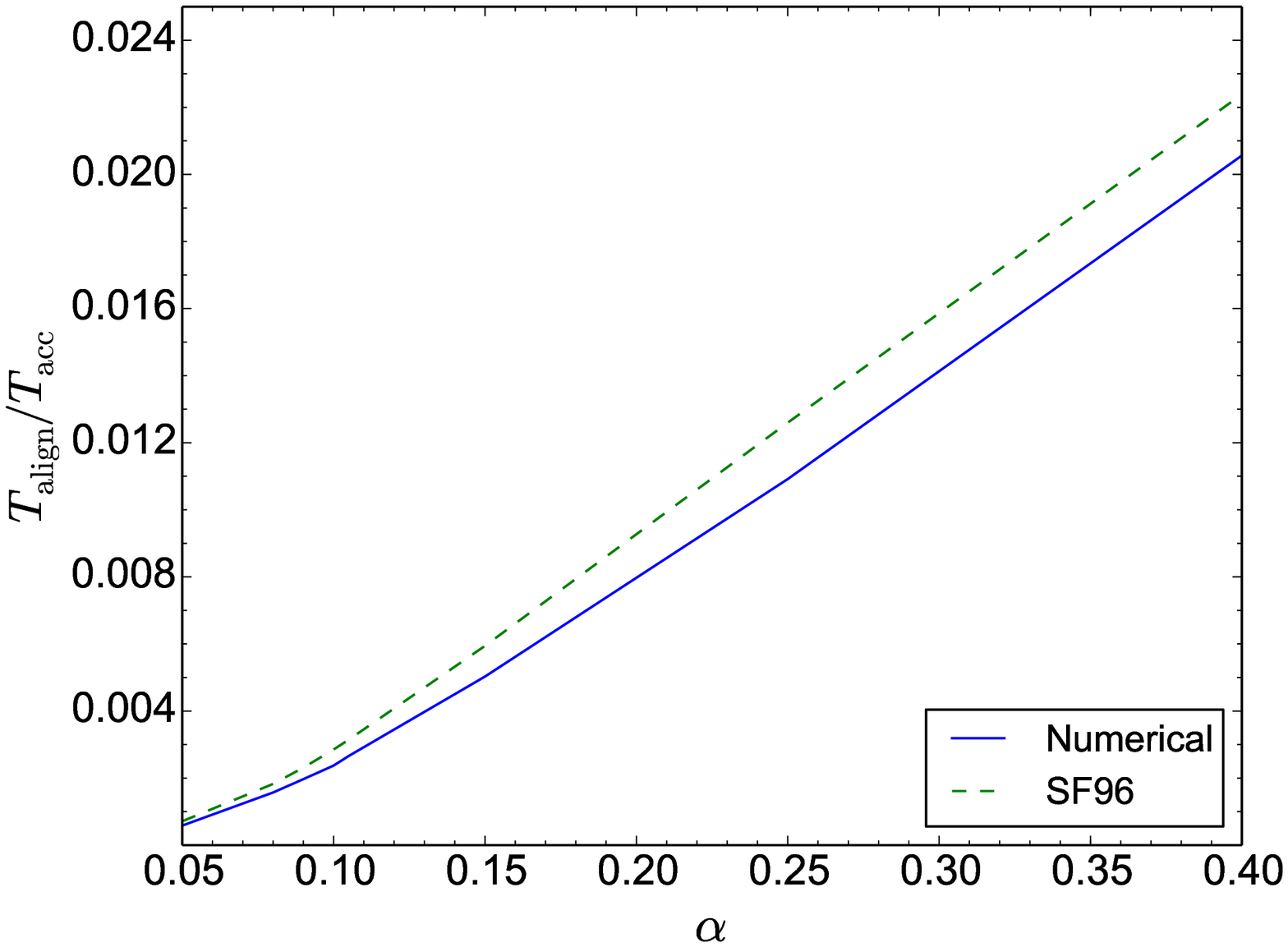}
\includegraphics[width=0.49\textwidth]{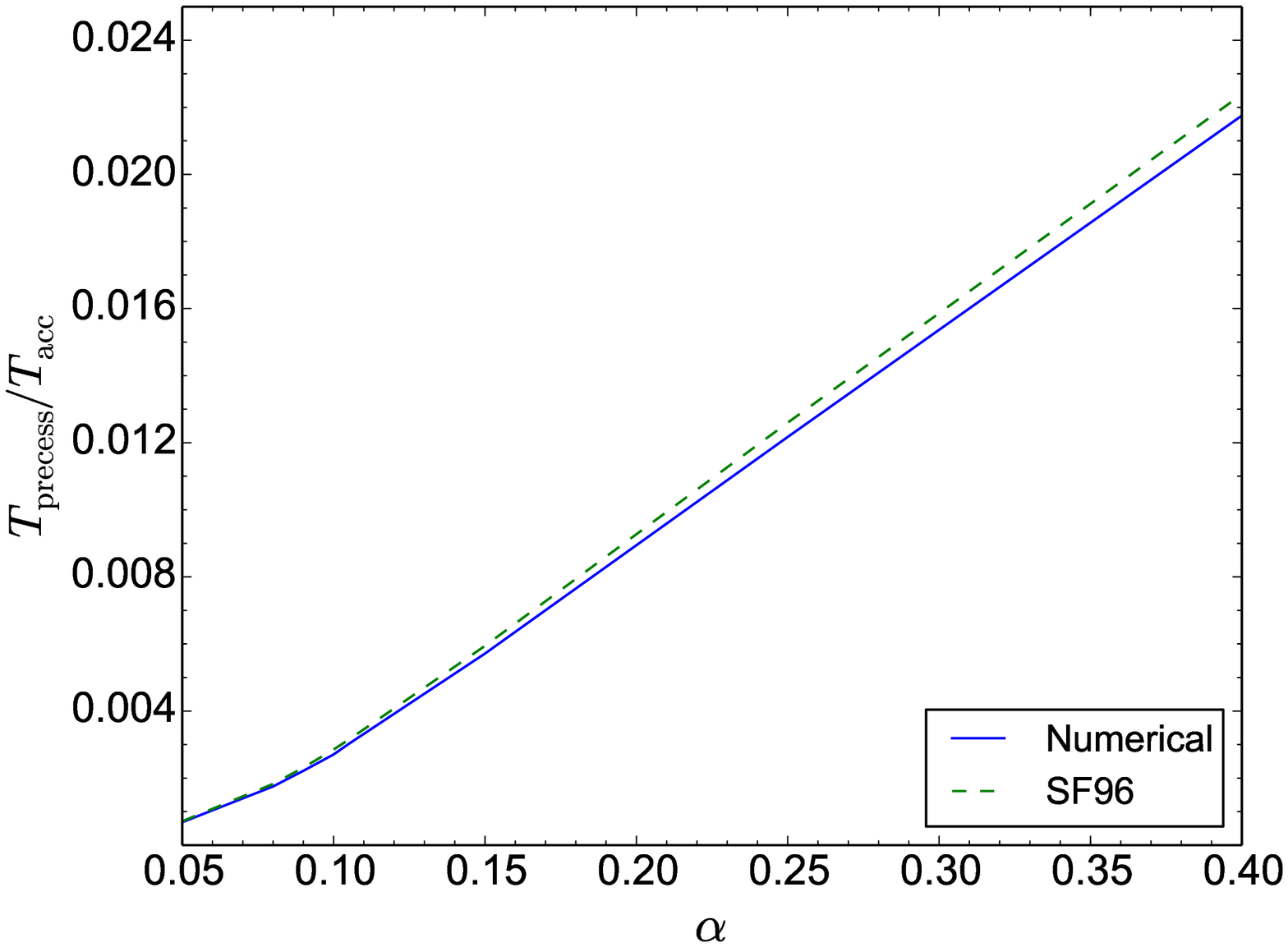}
\caption{\label{fg6}Alignment (Left panel) and precession (Right panel) time-scales for different values $\alpha$. Other parameter values are:  $\dot{M}=10^{-8}\ M_{\odot}\ \rm yr^{-1}=0.05\ \dot{M}_{\rm Edd}$ for $M=10\ M_{\odot}$, $a=0.2$, and $\nu_{20}=10^{15}\ \rm cm^2\ \rm s^{-1}$. Here, accretion time-scale, $T_{\rm acc}=\frac{M}{\dot{M}}$. The term `Numerical' signifies our numerically obtained result, and `SF96' denotes the result obtained by SF96 (see section \ref{per} for more details).\label{fig6}}
\end{figure*}

We also study the behaviour of the time-scales as functions of $\alpha$ in Fig. \ref{fg6}. We find that the deviation between our result and the result of SF96 increases with an increase in $\alpha$. This can be explained from the fact that, as one increases $\alpha$, $n$ also increases (see the equation (\ref{alpha}) and the expression of $n$ below equation (\ref{noeq})) and thus the term, which was ignored in SF96 (i.e., the term containing $C_1$ in the equation (\ref{warp}) or the term associated with $C_1$ in the equations (\ref{noeq}); see section 4.3 of BCB19), being proportional to $n$ also increases resulting in more deviation from the result of SF96. But the departure from the SF96 results in this case is small for the chosen range of $\alpha$ compared to the deviations discussed in previous paragraphs. This behaviour was also noted by \cite{Lodato}. 

Thus we find that, for the low viscosity value (when $M$ and $a$ are fixed) or the high Kerr parameter value  (when $M$ and $\nu_2$ are fixed), our results are in agreement with the SF96 results. But for the other end of parameter space, departure from SF96 results can become important. This behaviour in the context of radial distribution of the disc tilt angle was also noted earlier by BCB19, \cite{Lodato} as discussed above. 
\begin{figure*}
\centering
\includegraphics[width=0.49\textwidth]{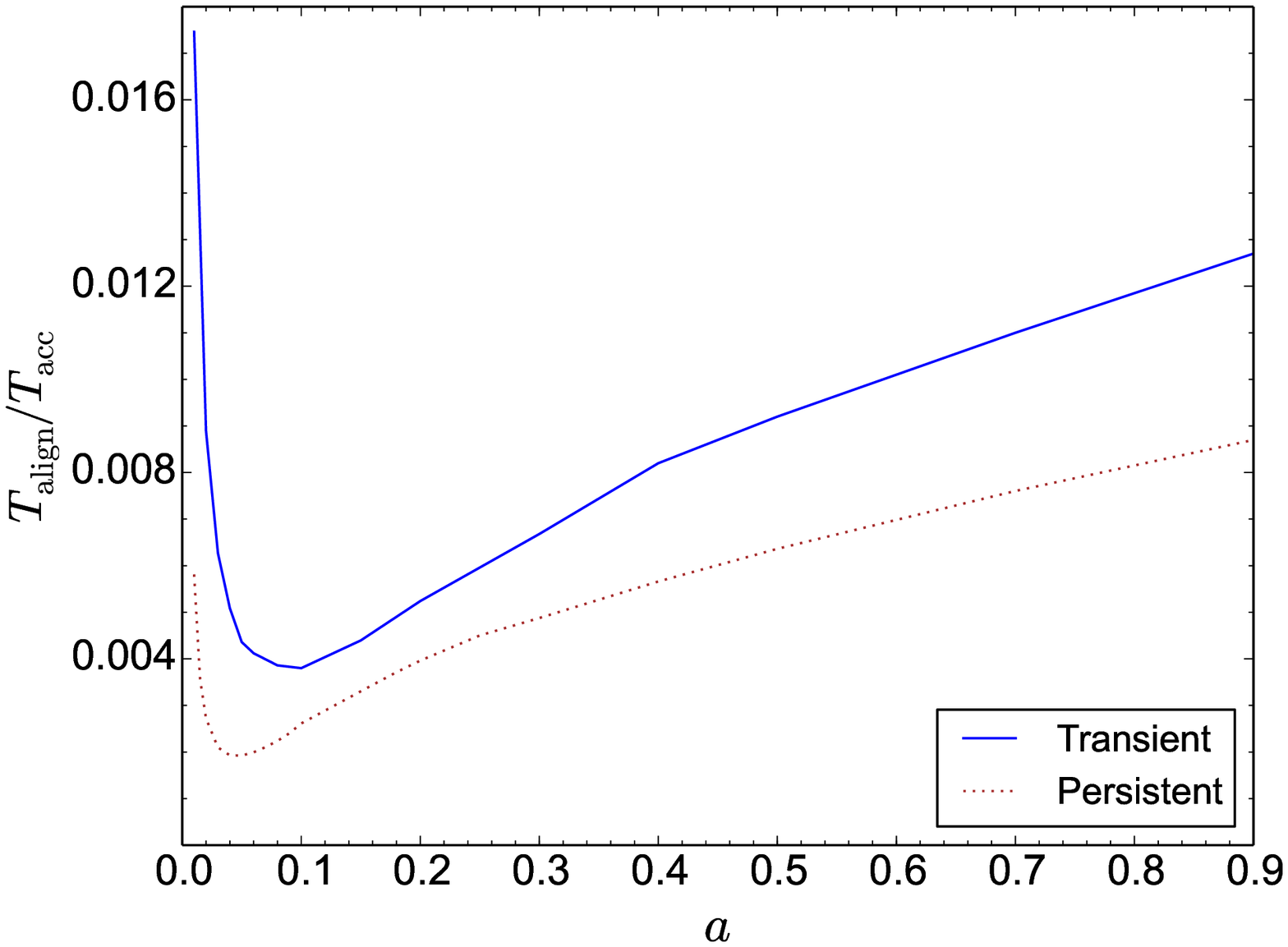}
\includegraphics[width=0.49\textwidth]{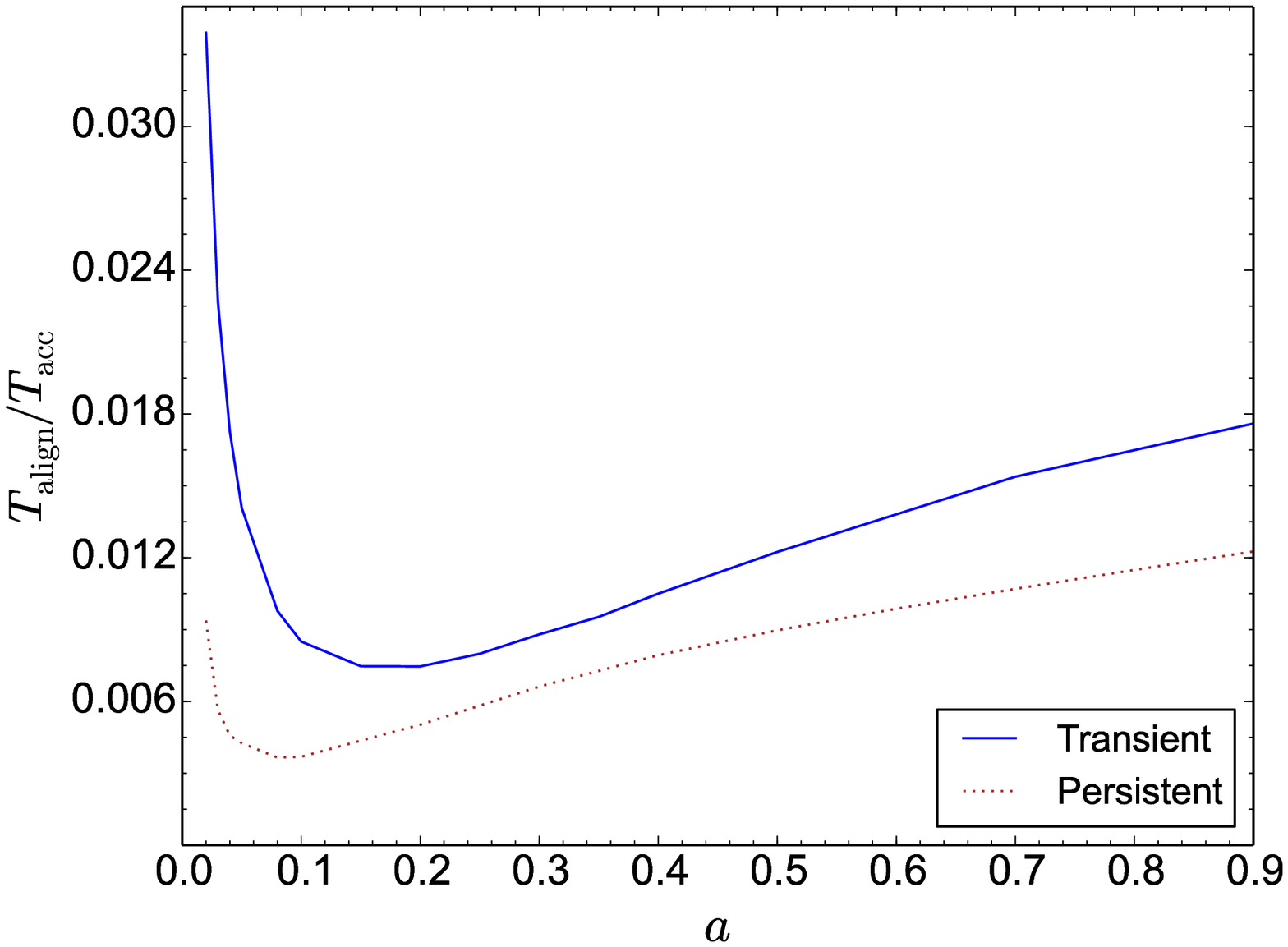}
\caption{\label{fg7}Alignment time-scales for different values of the Kerr parameter for the viscosities $\nu_{20}=5\times10^{14}\ \rm cm^2\ \rm s^{-1}$ (Left panel) and $\nu_{20}=10^{15}\ \rm cm^2\ \rm s^{-1}$ (Right panel). Other parameter values are:  $\dot{M}_{\rm av}=10^{-8}\ M_{\odot}\ \rm yr^{-1}=0.05\ \dot{M}_{\rm Edd}$ for $M=10\ M_{\odot}$, and $\alpha=0.15$. Here, accretion time-scale, $T_{\rm acc}=\frac{M}{\dot{M}}$, and the persistent accretion rate is same as the average accretion rate for the transient accretion (see section \ref{numtran} for more details). \label{fig7}}
\end{figure*}
\subsection{Numerical Results: Transient Accretion}\label{numtran}
\begin{figure*}
\centering
\includegraphics[width=0.49\textwidth]{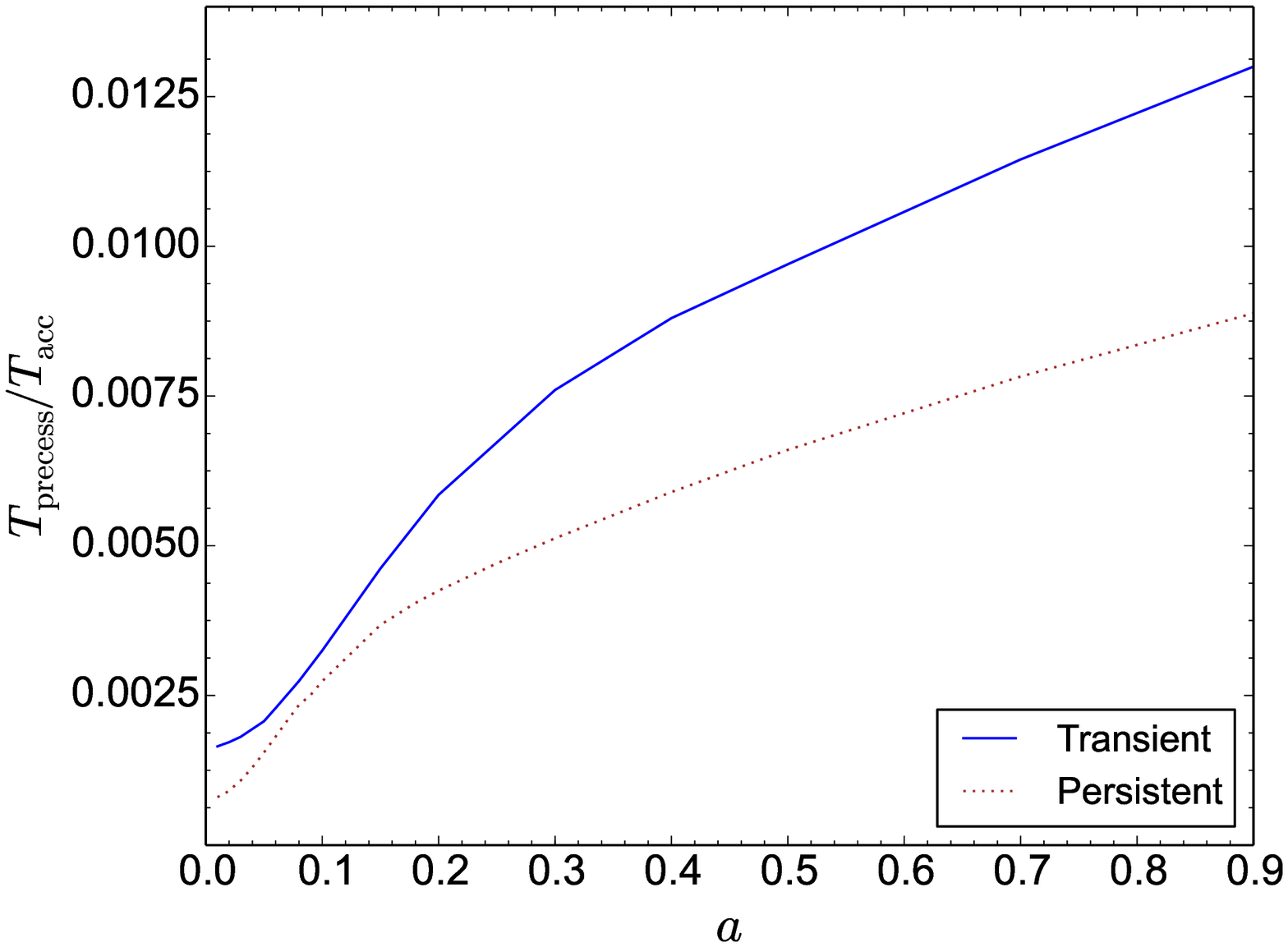}
\includegraphics[width=0.49\textwidth]{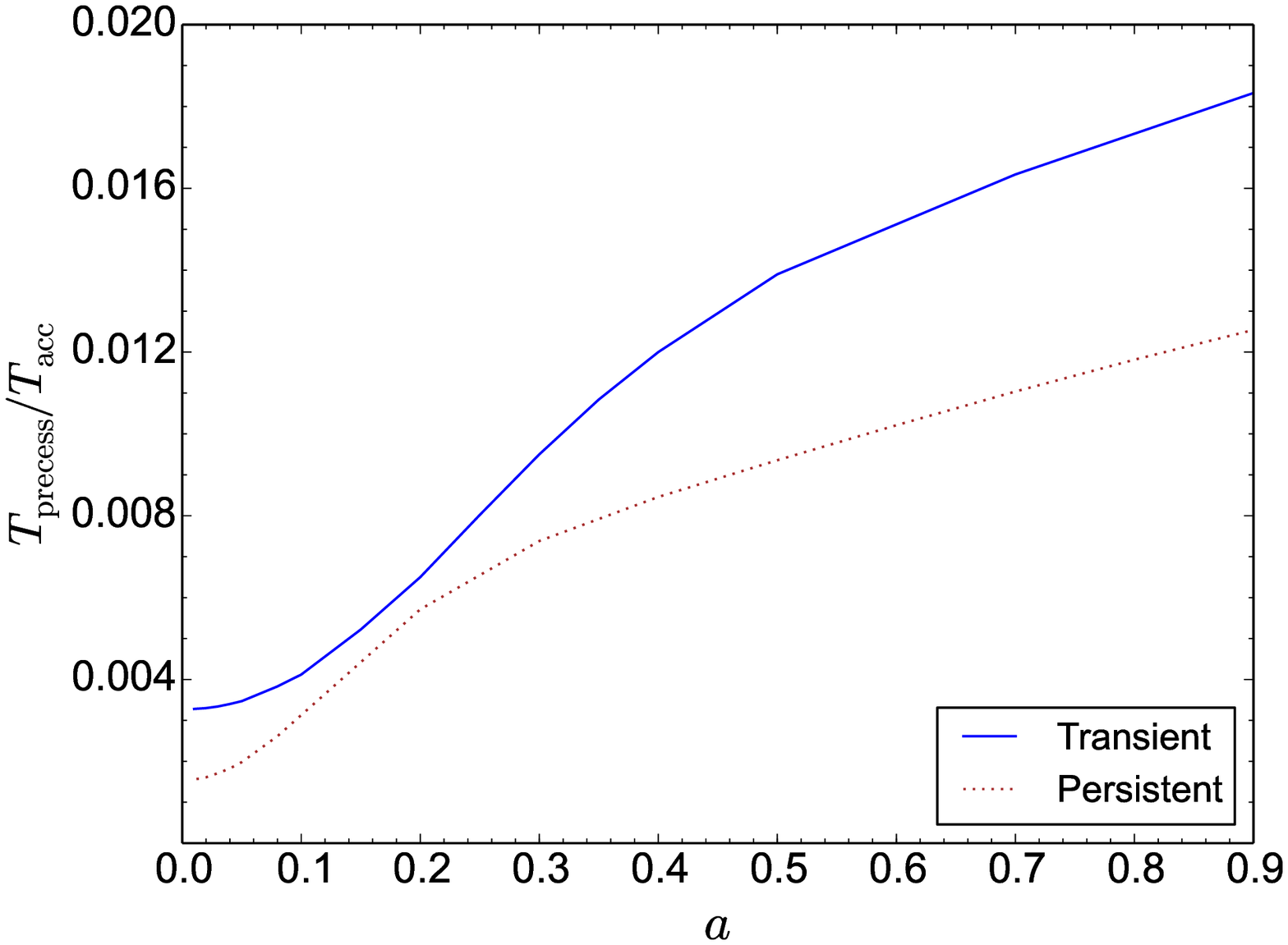}
\caption{\label{fg8}Precession time-scales for different values of the Kerr parameter for the viscosities $\nu_{20}=5\times10^{14}\ \rm cm^2\ \rm s^{-1}$ (Left panel) and $\nu_{20}=10^{15}\ \rm cm^2\ \rm s^{-1}$ (Right panel). Other parameter values are:  $\dot{M}_{\rm av}=10^{-8}\ M_{\odot} \rm yr^{-1}=0.05\ \dot{M}_{\rm Edd}$ for $M=10\ M_{\odot}$, and $\alpha=0.15$. Here, accretion time-scale, $T_{\rm acc}=\frac{M}{\dot{M}}$, and the persistent accretion rate is same as the average accretion rate for the transient accretion (see section \ref{numtran} for more details). \label{fig8}}
\end{figure*}
\begin{figure*}
\centering
\includegraphics[width=0.49\textwidth]{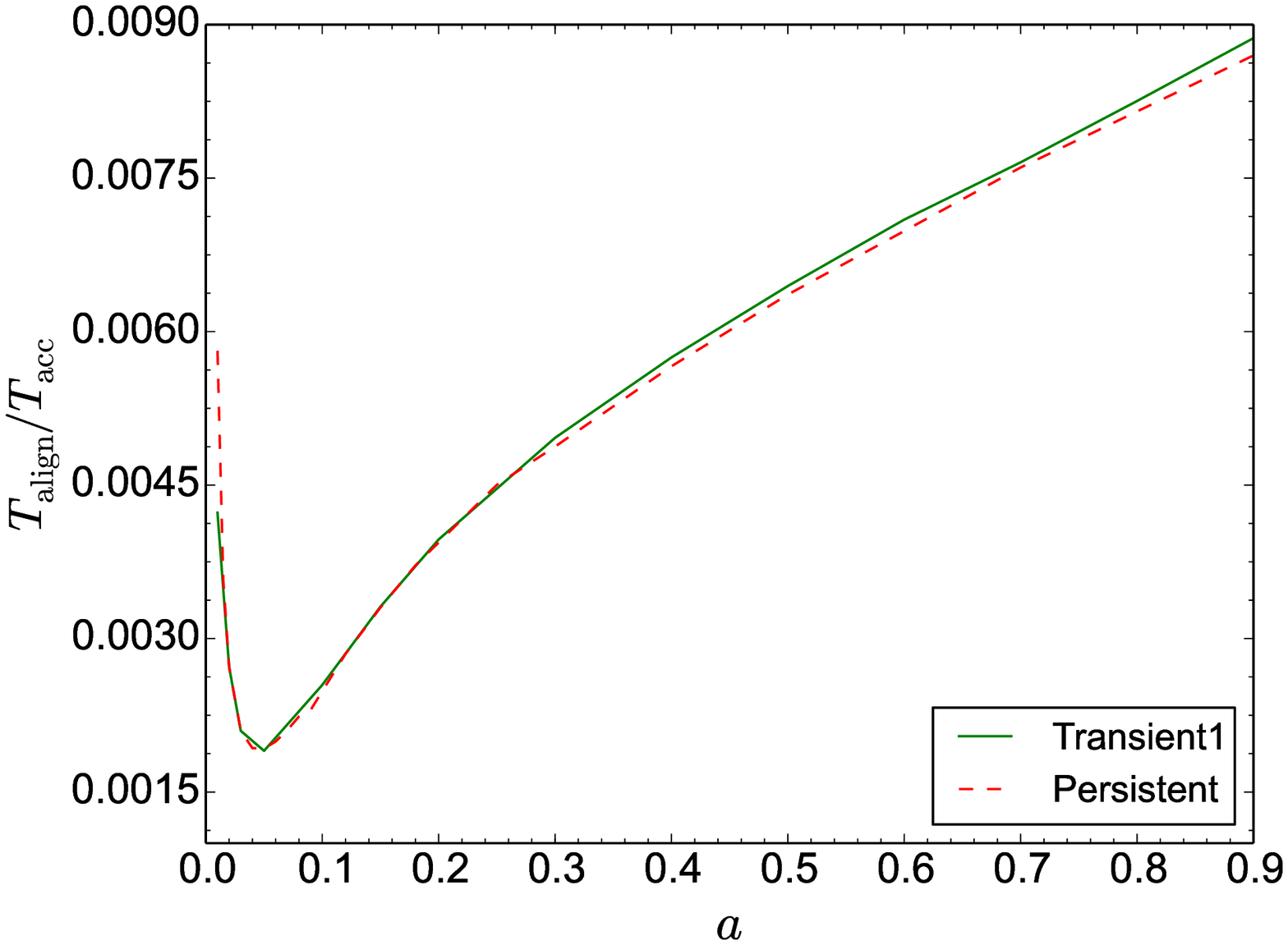}
\includegraphics[width=0.49\textwidth]{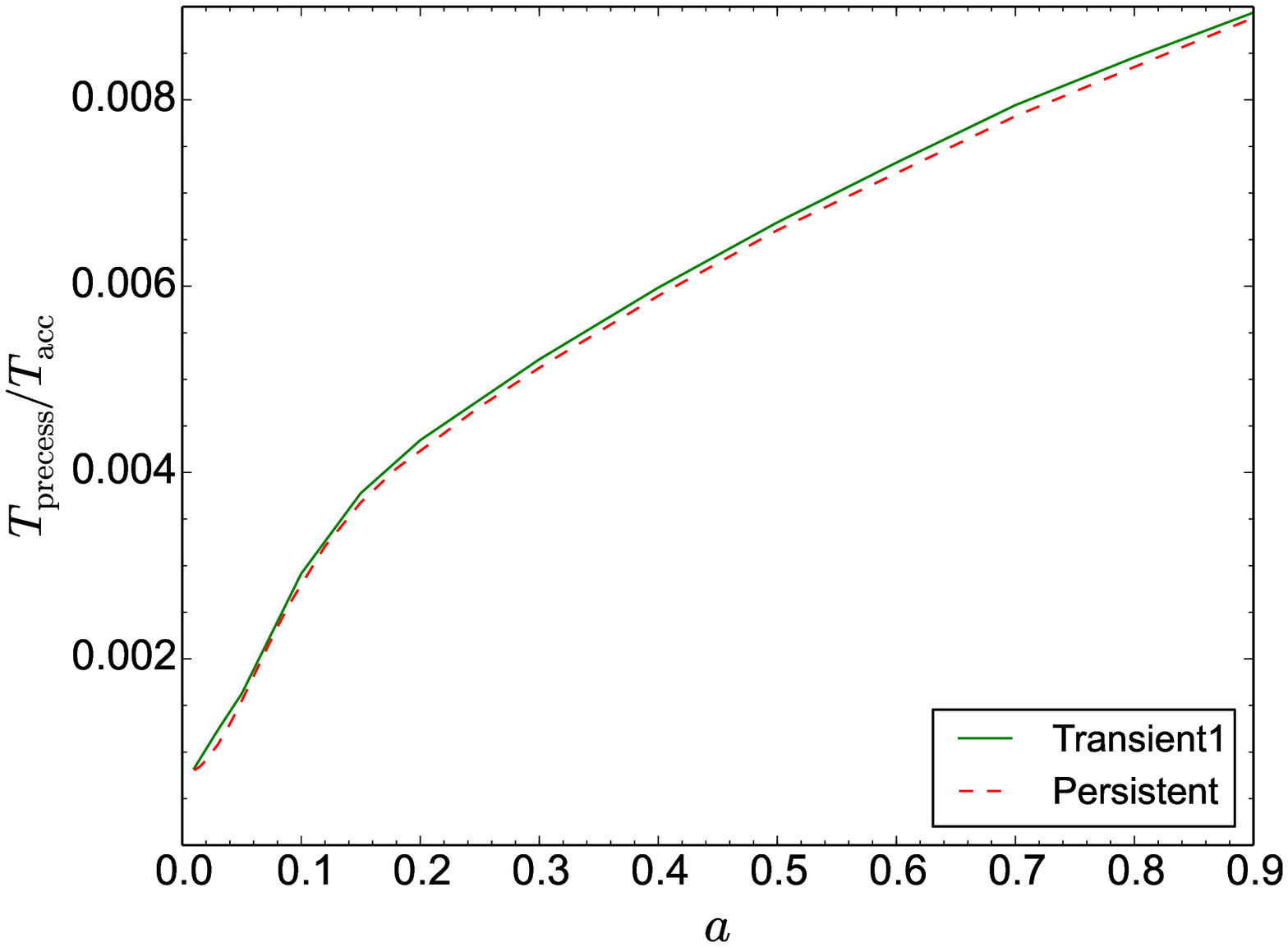}
\caption{\label{fg8a}Alignment (Left panel) and precession time-scales (Right panel) for different values of the Kerr parameter for the viscosity $\nu_{20}=5\times10^{14}\ \rm cm^2\ \rm s^{-1}$. Other parameter values are:  $\dot{M}_{\rm av}=10^{-8}\ M_{\odot}\ \rm yr^{-1}=0.05\ \dot{M}_{\rm Edd}$ for $M=10\ M_{\odot}$, and $\alpha=0.15$. Here, accretion time-scale, $T_{\rm acc}=\frac{M}{\dot{M}}$, and the term `Transient1' signifies the case of transient accretion when viscosity is independent of accretion rate. The persistent accretion rate is same as the average accretion rate for the transient accretion (see section \ref{numtran} for more details). \label{fig8a}}
\end{figure*}
We now discuss our results for transient accretion in comparison to what we have obtained for persistent accretion. For calculating the time-scales for transient accretion, we assume that the evolution of accretion rate during an outburst can be modelled by a triangular profile (see Fig. \ref{fg0}) as discussed in section \ref{tran}, and all the outbursts during the alignment process are described by the same profile of accretion rate. For modelling the triangular profile of accretion rate during the outburst phase, we consider the decay time (i.e., the time interval from the maximum accretion rate to the quiescent phase) in the outburst interval twice the rise time (i.e., the time interval from the quiescent phase to the maximum accretion rate). We also check our results when the decay time is thrice the rise time, and find no change in our qualitative conclusion. We also assume duty cycle($f$) $=0.1$, and the peak X-ray luminosity to be $L_{\rm Edd}$ for the illustration purpose (unless otherwise stated). For a $10\ M_{\odot}$ stellar mass black hole, which we choose in this section, the Eddington accretion rate is $\sim2\times 10^{-7} M_{\odot}\ \rm yr^{-1}$.

Alignment and precession time-scales in the transient accretion qualitatively 
follow the same trend as found in the case of persistent accretion as functions of the Kerr parameter. But they (i.e., time-scales in the transient case) are longer than the same obtained for the case of persistent accretion for the same average accretion rate (Figs. \ref{fg7} and \ref{fg8}). Such an enhancement can be understood by analysing the behaviour of the time-scales as functions of the viscosity parameter $\nu_2$.

As we see from the equation (\ref{num}), the time-scales depend on $C_1$ (see equation (\ref{C1n})) and the integrals $\mathcal{I}_1$ and $\mathcal{I}_2$. The contribution of the integrals $\mathcal{I}_1$ and $\mathcal{I}_2$ mainly depend on the radial profiles of $l_y$ and $l_x$ respectively, and the viscosity parameter $\nu_2$ strongly decides the radial profiles of these disc tilt vector components through warped disc equations (\ref{warp1},\ref{warp2}). As $\nu_2$ takes a higher value, the viscous torque in the plane of the disc ($\nu_{2}$ is the viscosity parameter associated with this torque; see section \ref{for1} for more details) hinders the alignment process more. As a result of which both the tilt vector radial profiles contribute more giving larger integral values. Thus the values of the integrals $\mathcal{I}_1$ and $\mathcal{I}_2$ increase with the increase in $\nu_2$. But, as $\nu_{2}$ increases, the quantity $C_1$ (see equation (\ref{C1n})) falls faster than the increase in the values of the integrals $\mathcal{I}_1$ and $\mathcal{I}_2$. Therefore, the time-scales increase with the increase of $\nu_2$ (see Fig. (\ref{fig6})). For transient accretion, when viscosity is independent of the accretion rate, the alignment and precession time-scales almost remain identical to that obtained for the persistent accretion with the same average accretion rate (see Fig. \ref{fg8a} ; only for this figure in this work we consider the viscosities to be independent of accretion rate).
But in our case of transient accretion, the viscosities are dependent upon the accretion rate and the accretion rate gets greatly enhanced during the outburst phase. Thus the time-scales take a higher value than the case when viscosity is accretion rate independent (as considered in \cite{Maccarone}; see section \ref{tran} for more details).    
\begin{figure*}
\centering
\includegraphics[width=0.49\textwidth]{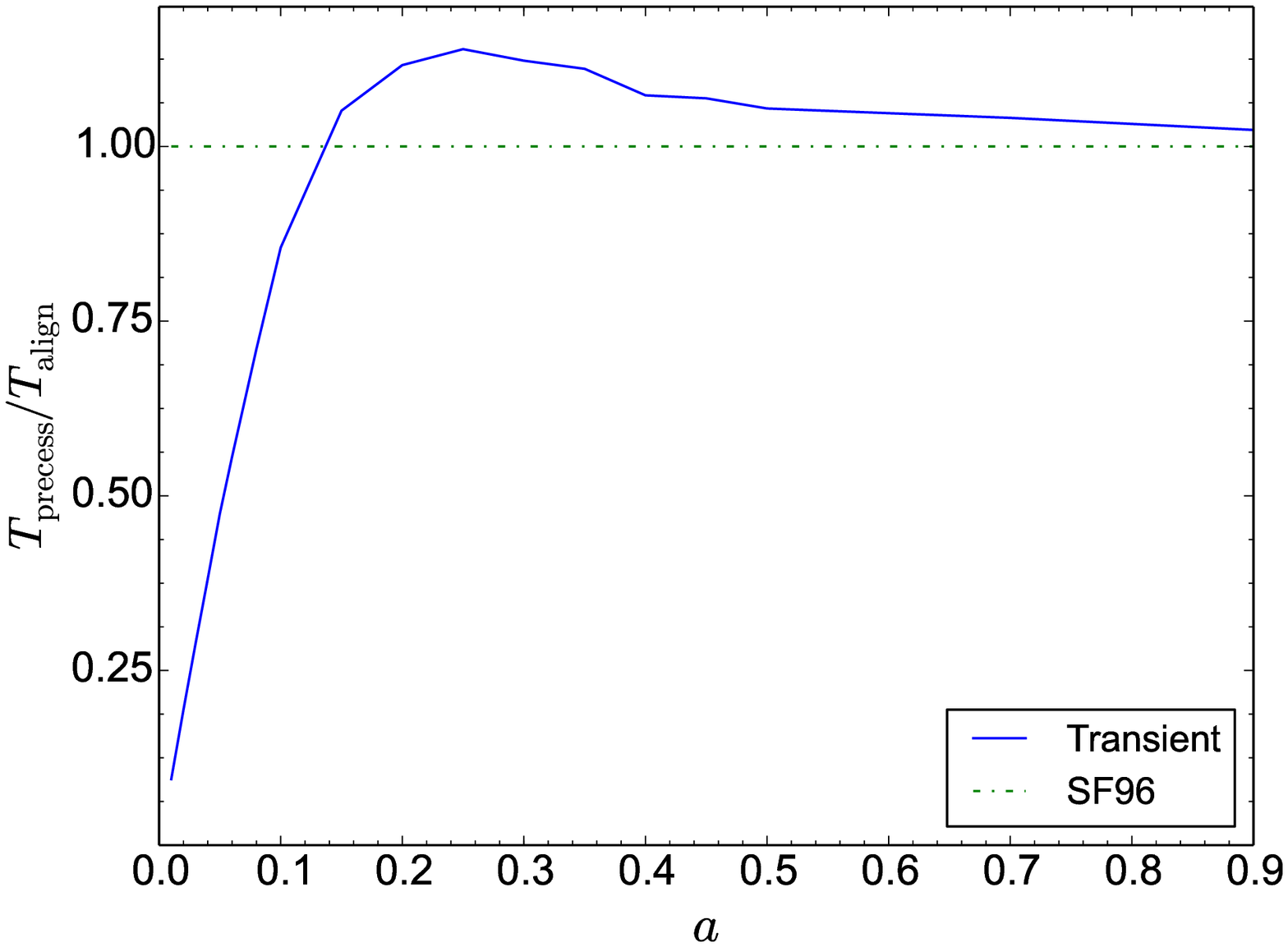}
\includegraphics[width=0.49\textwidth]{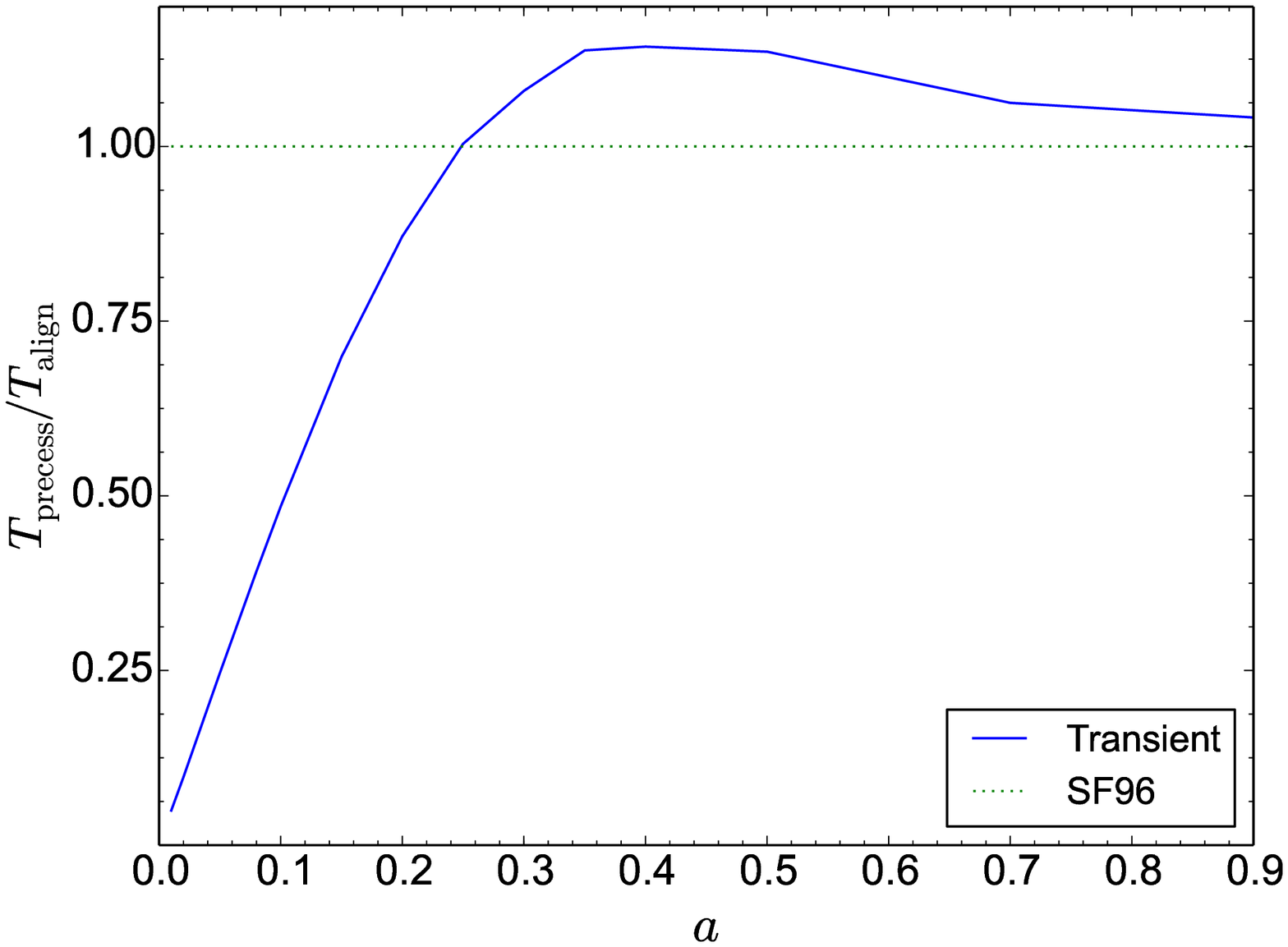}
\caption{\label{fg9}Ratio between precession and alignment time-scales in transient accretion for different values of the Kerr parameter for the viscosities $\nu_{20}=5\times10^{14}\ \rm cm^2\ \rm s^{-1}$ (Left panel) and $\nu_{20}=10^{15}\ \rm cm^2\ \rm s^{-1}$ (Right panel). Other parameter values are: $\dot{M}_{\rm av}=10^{-8}\ M_{\odot}\ \rm yr^{-1}=0.05\ \dot{M}_{\rm Edd}$ for $M=10\ M_{\odot}$, and $\alpha=0.15$ (see section \ref{numtran} for more details). \label{fig9}}
\end{figure*}
\begin{figure*}
\centering
\includegraphics[width=0.49\textwidth]{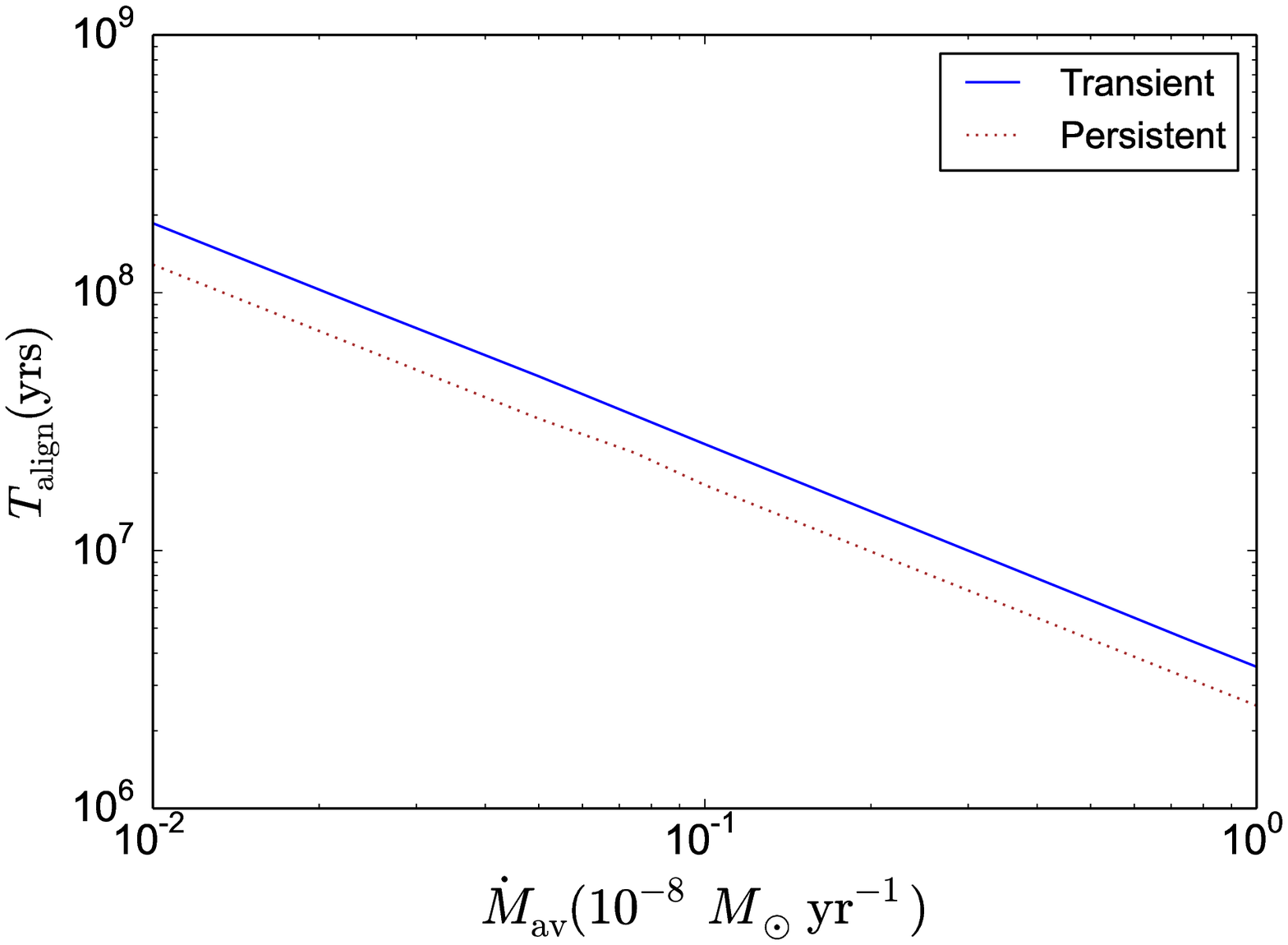}
\includegraphics[width=0.49\textwidth]{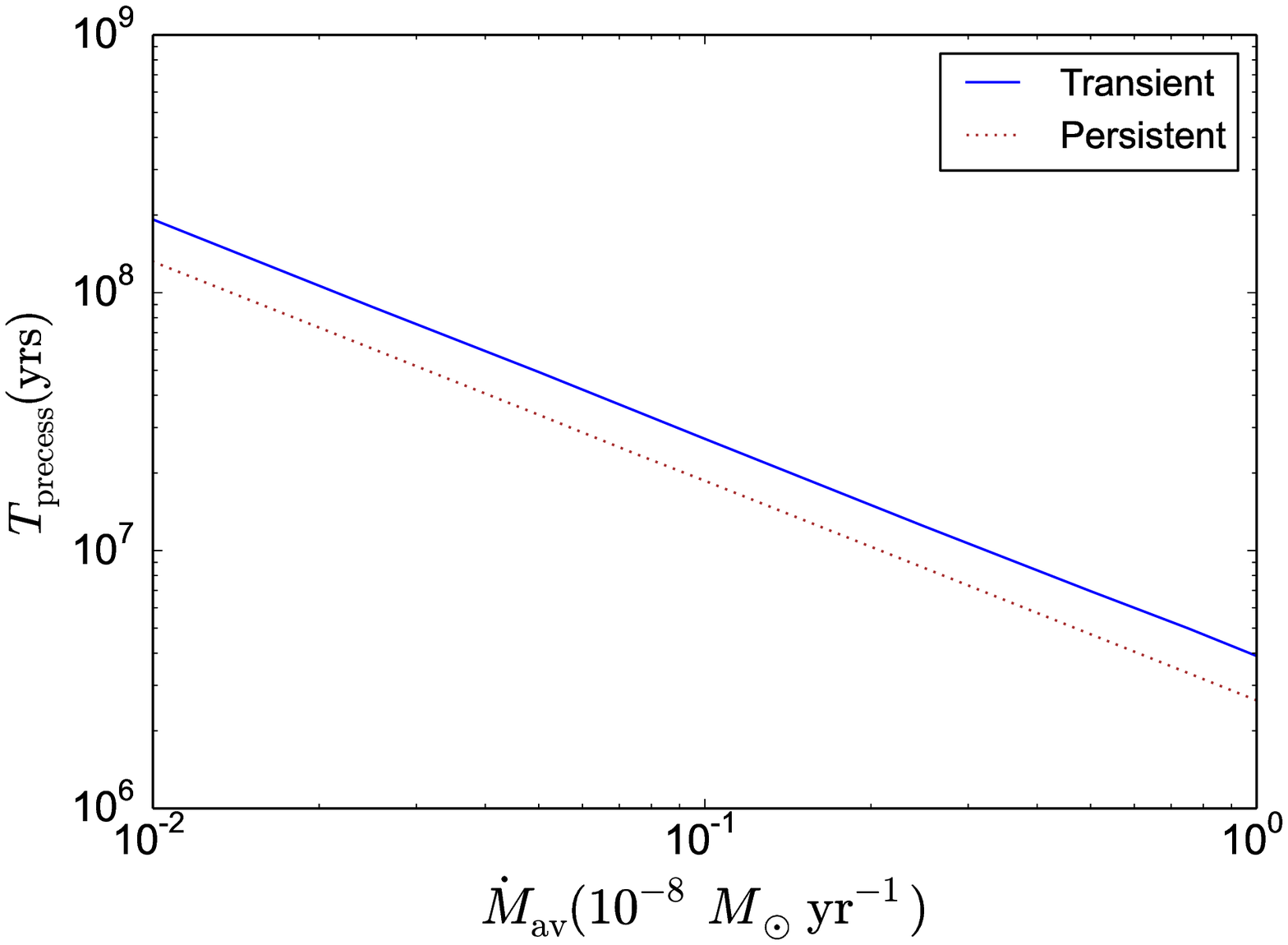}
\caption{\label{fg10}Alignment (Left panel) and Precession time-scales (Right panel) for different values of average accretion rates. Other parameter values are: $M=10\ M_{\odot}$, $\alpha=0.15$, $\nu_{20}=5\times10^{14}$ $\rm cm^2\ \rm s^{-1}$ and $a=0.4$. Here, the persistent accretion rate is same as the average accretion rate for the transient accretion (see section \ref{numtran} for more details). \label{fig10}}
\end{figure*}
In the case of transient accretion too, there exists a threshold value of the Kerr parameter below which the alignment time-scale decreases with the increase in Kerr parameter, and it also depends on the viscosity $\nu_2$ (Fig. \ref{fg7}) and mass of the black hole (the reason behind such a behaviour is similar to the case of persistent accretion). We also find that the difference between the time-scales' value computed for transient and persistent accretions is higher for higher values of the Kerr parameter beyond this threshold (Figs. \ref{fg7} and \ref{fg8}). Similar to the case of persistent accretion, the ratio between precession time-scale and alignment time-scale is different from unity unlike the SF96 case, and slowly moves toward unity for higher values of the Kerr parameter (Fig. \ref{fg9}). We also compare the accretion rate dependence (i.e., average accretion rate, see equation (\ref{avg})) of the time-scales for both the cases, and find the time-scales for transient accretion are almost 1.5 times longer than the case of persistent accretion for the same average accretion rate (Fig. \ref{fg10}; we vary the duty cycle in these panels). Like the case of persistent accretion, the contribution of the inner disc for transient accretion (i.e., deviation from the behaviour of the time-scales as suggested by SF96) can also be understood from the ratio between the quantities $R_{\rm in}$ and $R_{w}$ (see section (\ref{per})).

We note here that during the quiescent state, where accretion rate is very low compared to the same in the outburst phase, the accretion process may not be described by a thin Keplerian viscous disc \citep{Menou}. Rather accretion phenomenon during the quiescence is usually delineated by the advection dominated accretion flow \citep{Yi,Menou} for which the disc thickness $H$ is roughly equal to $R$ \citep{Maccarone}. Therefore, our formalism, which assumes the disc to be thin, viscous and Keplerian, may not hold for the quiescent phase. But, because of very low accretion rates during the quiescent phase, even if the accretion flow is more efficient in transferring warps than in the outburst phase, and even though such a phase lasts for a longer period of time, the diffusion of warps in the outburst phase may dominate the scenario \citep{Maccarone}. This also justifies our assumption of ignoring the change of black hole spin direction during the quiescent state as discussed in the section \ref{tran}.

\section{Application to black hole LMXBs: Determination of alignment time}\label{lmxb}
The majority of the Galactic black hole X-ray binaries (BHXBs) exhibit transient behaviour. Most transient BHXBs are low mass X-ray binary (LMXB) systems \citep{Tetarenko}, composed of a black hole accreting from a less massive star (spectral type A or later; \cite{Liu}), in which mass transfer occurs episodically through Roche-lobe overflow. As mentioned earlier in section \ref{tran}, such systems can be characterized by two distinct phases of activity, outburst and quiescent phases, in which most of the matter accretion happens during the outburst phase triggered possibly by the thermo-viscous disc instability \citep{Yu}. In this work, we calculate the alignment time $t_{\rm a}$ (one should note that it is different from the alignment time-scale $T_{\rm align}$), i.e., the time the system takes in reducing the misalignment between the black hole spin and outer disc angular momentum direction to $<1\%$ of the initial value, for three black holes in LMXBs 4U 1543--475, H 1743--322 and XTE J1550--564. Besides, we calculate the alignment time $t_{\rm a}$ of a black hole in a typical Galactic LMXB using the average properties of an outburst (as observed in the {\it{RXTE}} era) in such a system as determined by \cite{Yu}. We also discuss the implications of our result in the context of spin measurement of black hole using the continuum X-ray spectral fitting method (see section \ref{intro}).  

For calculating the alignment time of black holes in the above mentioned three LMXBs and also in typical Galactic black hole transient LMXBs, we assume accretion history of these systems can be described by a series of identical triangular profile of accretion rate during outbursts as shown in Fig. \ref{fg0}. For estimating the shape of the triangular profile for a black hole, we consider outburst light curve(s) for that black hole as observed with {\it{RXTE}}/ASM, and fit the accretion rate profile during the rise time (the time the system takes to rise to the peak X-ray flux from $1\%$ of the peak flux) and the decay time (the time the system takes to decay to $1\%$ of the peak flux from the peak X-ray flux) of the outburst for that black hole by a straight line. We also need the peak accretion rate for uniquely defining the triangular profile, and it can be obtained from the information of peak X-ray flux and distance to the system.
Once we model the evolution of accretion rate during an outburst by a triangular profile from the observed light curve of that outburst, we assume this outburst profile as a prototype for all the outbursts happened during the alignment time of the black hole. We also approximate the evolution of the accretion rate during the decay time by an exponential function for the black hole in LMXB 4U 1543--475, and find that the result remains unaffected within $5\%$. We choose $\alpha=0.1$, and the values of the other parameters as specified in the section $\ref{param}$ (we consider both the viscosities to be accretion rate dependent throughout this section as defined by the equation (\ref{vis})). The numerical scheme we consider for calculating the alignment time is described in the section \ref{tran}. Here, we also intend to check whether the assumption of alignment of the black hole's spin with the disc outer edge is consistent with the spin value of the black hole as estimated by the continuum X-ray spectral fitting method. Thus, we consider the spin value of the black hole as determined by this method for calculating the alignment time of that black hole. We have seen previously that the warp radius $R_w$ (equation (\ref{warprad})) plays an important role in the alignment time-scale calculation in our as well as earlier works. We find that our chosen range of parameter values for estimating the alignment time for each black hole LMXB, lead to a large range of the warp radius indicating the robustness of our results. 
\subsection{Alignment time: 4U 1543--475}\label{4u}
\begin{figure}
\begin{center}
\includegraphics[width=0.49\textwidth]{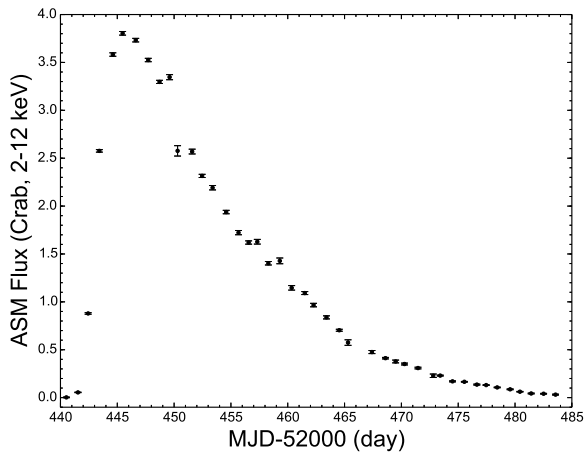}
\caption{\label{fg11}Light curve (with one day bin) of the black hole LMXB 4U 1543--475 during the 2002 outburst in the range 2-12 keV as observed with {\it{RXTE}}/ASM (see section \ref{4u}).\label{fig11}}
\end{center}
\end{figure}
The recurrent X-ray transient 4U 1543--475 was discovered in $1971$ \citep{Matilsky}, and went into outbursts in 1983 \citep{Kitamoto}, 1992 \citep{Harmon} and 2002 \citep{Park}. This black hole LMXB is composed of a black hole of mass $M=9.4\pm1 M_{\odot}$ \citep{Orsoz}, and a secondary companion of mass $2.7\pm 1 M_{\odot}$ \citep{Orsoz} of spectral type A2V \citep{Remillard}. Besides, the distance to the system was determined to be $D=7.5\pm 1.0\ \rm kpc$ \citep{Orsoz}. 
The spin of the black hole was estimated to be $0.75-0.85$ using the continuum X-ray spectral fitting method by \cite{Shafee}. 

We calculate the alignment time of this black hole by assuming the 2002 accretion rate outburst profile as the prototype for all the outbursts (see the light curve in Fig. \ref{fig11}), and model this outburst by a triangular profile as mentioned previously. We use the {\it{RXTE}}/ASM observation between MJD 52439--52483 (the X-ray flux remained within $1\%$ of the peak flux during this period) for modelling the 2002 outburst, in which the peak X-ray flux rose to $3.8$ Crab (2-12 kev) on MJD 52445 \citep{Yu}. The 2002 outburst exhibited the characteristic transitions in the Hardness-Intensity (HID) diagram \citep{Park}, i.e., `q' shape or turtle-head profile in the HID diagram \citep{Remillard}, and the system mostly stayed in the soft state during the above mentioned period. From the information of the distance (we choose $7.5\ \rm kpc$) and mass (we choose $9.4\ M_{\odot}$), one can estimate the peak accretion rate during the outburst, which roughly comes around $0.46\ \dot{M}_{\rm Edd}$ for this black hole. We take the spin of the black hole to be $a=0.8$, and the duty cycle of the outburst $\sim 0.02$ \citep{Tetarenko}. Using all these information, we find the alignment time $t_{\rm a}$ in the range $\sim 0.1-0.3\ \rm Gyr$ for the range of viscosity component $\nu_{20}$ (correspondingly a range of $\sim 17-640\ R_g$ for $R_w$; see equation (\ref{warprad}) for the definition of $R_w$) discussed in the section \ref{param}. The age of this system was predicted to be $0.1-0.6\ \rm Gyr$ \citep{Fragos}, and thus the spin of the black hole of the transient source 4U 1543--322 could remain misaligned with the outer disc. 

The spin of this black hole was also estimated from the iron K$\alpha$ line method by \cite{Miller}, and they found a lower value of spin, $0.3\pm0.1$ for this black hole. This discrepancy was shown to arise due to the assumption of alignment of the black hole spin with outer disc angular momentum in the continuum X-ray spectral fitting method, and the black hole was predicted to be slightly misaligned (by $\sim 12^{\circ}$; \cite{Morningstar}). Thus, our conclusion for this black hole regarding the misalignment is consistent with the prediction made by \cite{Morningstar}. 
\subsection{Alignment time: H 1743--322}\label{h1}
\begin{figure*}
\centering
\includegraphics[width=0.49\textwidth]{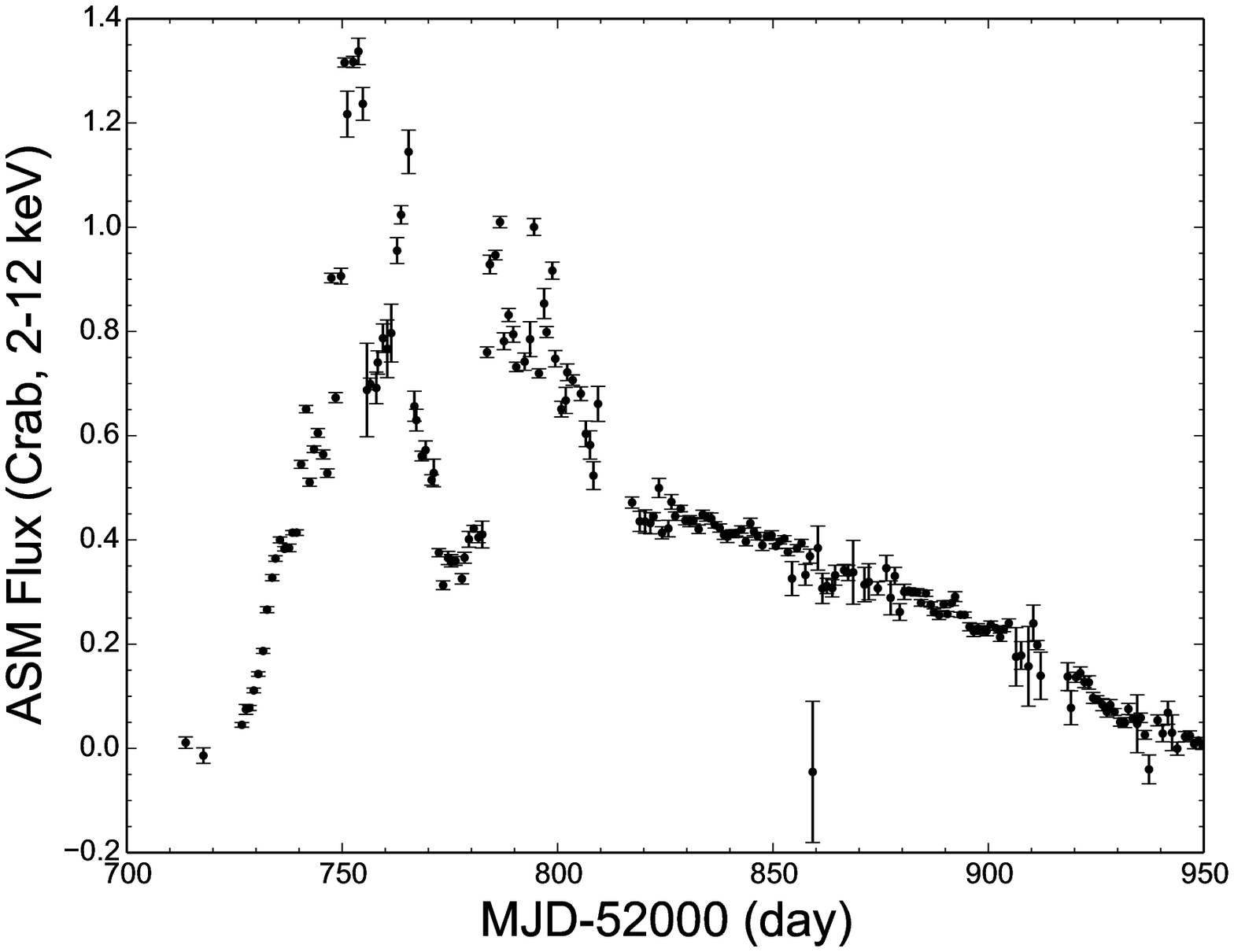}
\includegraphics[width=0.49\textwidth]{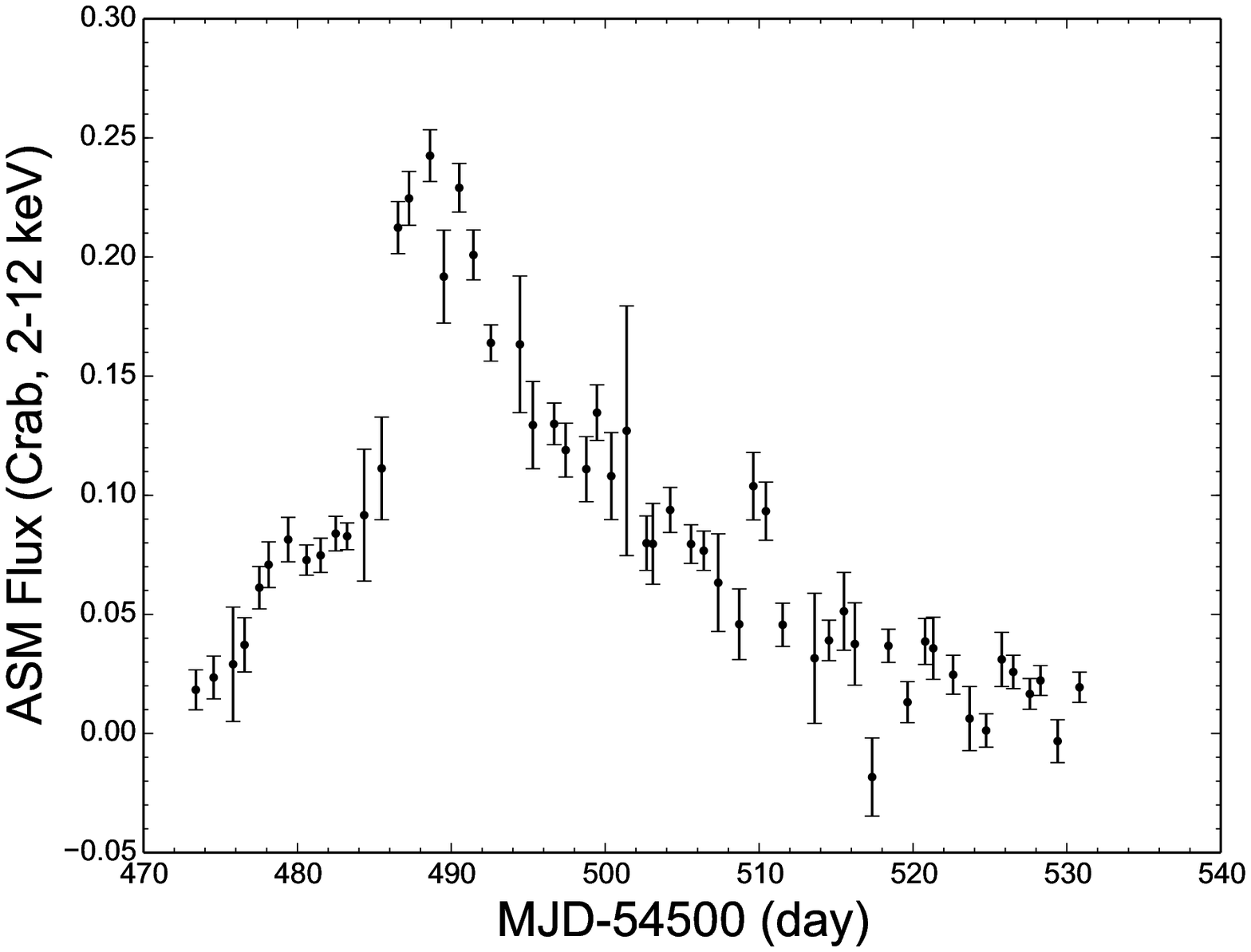}
\caption{\label{fg12}Light curves (with one day bin) of the black hole LMXB H 1743--322 during the 2003 (Left panel) and 2009 (Right panel) outbursts in the range 2-12 keV as observed with {\it{RXTE}}/ASM (see section \ref{h1}).\label{fig12}}
\end{figure*}
The X-ray transient H 1743--322 (also a microquasar) was discovered in 1977 with the Ariel V all sky monitor \citep{Kaluzeinski} and HEAO I satellite \citep{Doxsey}. This black hole LMXB has undergone six outbursts since 2003 with the soft disc dominated state (considering up to 2015 May; \cite{Tetarenko}), which we use for calculating the duty cycle (which is roughly $\sim0.07$). 
The mass of this black hole was determined to be $13.3\pm 3.2\ M_{\odot}$ by \cite{Titarchuk} using 2003 {\it{RXTE}}/PCA outburst data, and $11.2^{+1.65}_{-1.96}\ M_{\odot}$ by \cite{Molla} using 2010 and 2011 {\it{RXTE}}/PCA outburst data. But the mass of the secondary companion star is not known for this transient system. Besides the spin of the black hole was estimated to be $0.2\pm0.3$ using the continuum X-ray spectral fitting method by \cite{Steiner2}, and the source distance was determined to be $8.5\pm0.8\ \rm kpc$ \citep{Steiner2}.

In this work, we separately consider both the 2003 (MJD 52713--52950) and 2009 (MJD 54973--55031) outbursts as observed with {\it{RXTE}} (the X-ray flux remained within $1\%$ of the peak flux during both of these periods) for determining the alignment time of the black hole, since the outburst accretion rate profiles were significantly different in them (see the light curves in Fig. \ref{fg12}). As can be seen in Fig. \ref{fg12}, the 2003 outburst was much more intense than the one in 2009, and the X-ray flux rose to $1.39$ Crab (2-12 keV) on MJD 52753 for the former, while the flux rose to $0.25$ Crab on MJD 54988 for the latter. Thus, if one considers the 2003 outburst profile as the prototype for all the outbursts, one obtains a shorter alignment time than the one obtained considering the 2009 outburst as the prototype. Here, we note that both of these outbursts are dominated by soft disc dominated state, and the X-ray radiation mostly stayed in soft state during the above mentioned periods . For calculating $t_{\rm a}$, we adopt $M=12\ M_{\odot}$ and distance $=8.5\ \rm kpc$. One can find the peak X-ray luminosity from the peak X-ray flux and distance, and it comes around $0.168\ L_{\rm Edd}$ for the 2003 outburst, and $0.042\ L_{\rm Edd}$ for the 2009 outburst. Using all these information, the alignment time $t_{\rm a}$ comes around $0.05-0.1\ \rm Gyr$ if one considers the 2003 outburst profile as the prototype for all the outbursts, and $t_{\rm a}$ lies in the range $0.09-0.32\ \rm Gyr$ as one assumes the 2009 outburst profile as the prototype, for the range of viscosity component $\nu_{20}$ mentioned in section \ref{param} (corresponding range of $R_w$ is $8-212$ $R_g$; see the equation (\ref{warprad}) for the definition of $R_w$). We note that considering two different outbursts as prototypes, and combining the results from both make our analysis robust.

The inner accretion disc around this black hole could be misaligned with spin direction of the black hole \citep{Ingram} indicating the black hole spin to be tilted with the outer disc. Hence, there is a possibility that this black hole may not have got sufficient time to align itself with the outer disc. Therefore, while the age of this system is not known due to the lack of knowledge about the companion, if we consider that the black hole spin is misaligned with the outer disc, the age of the LMXB should be roughly less than $0.32\ \rm Gyr$. 
\subsection{Alignment time: XTE J1550--564}\label{xte}
\begin{figure*}
\centering
\includegraphics[width=0.49\textwidth]{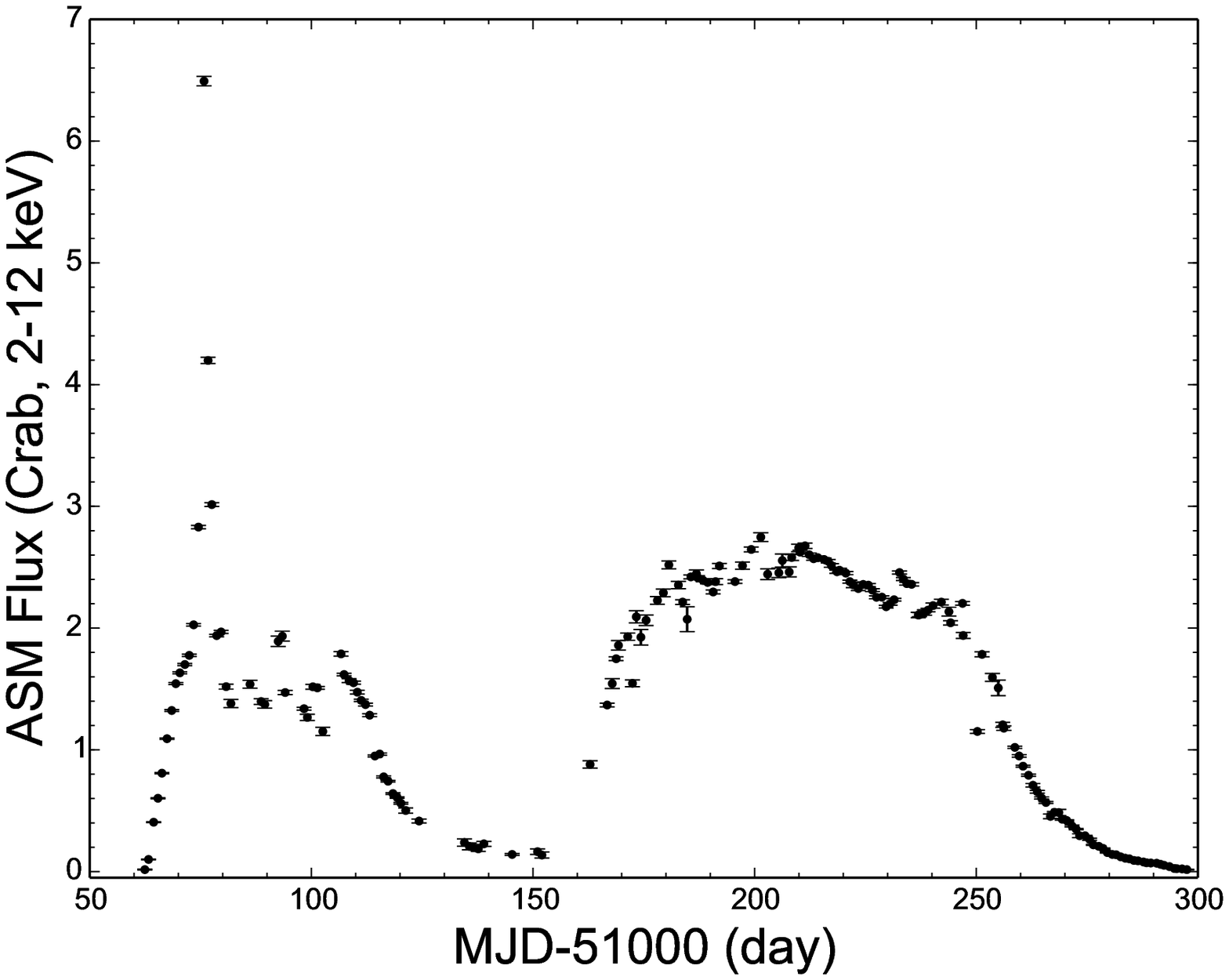}
\includegraphics[width=0.49\textwidth]{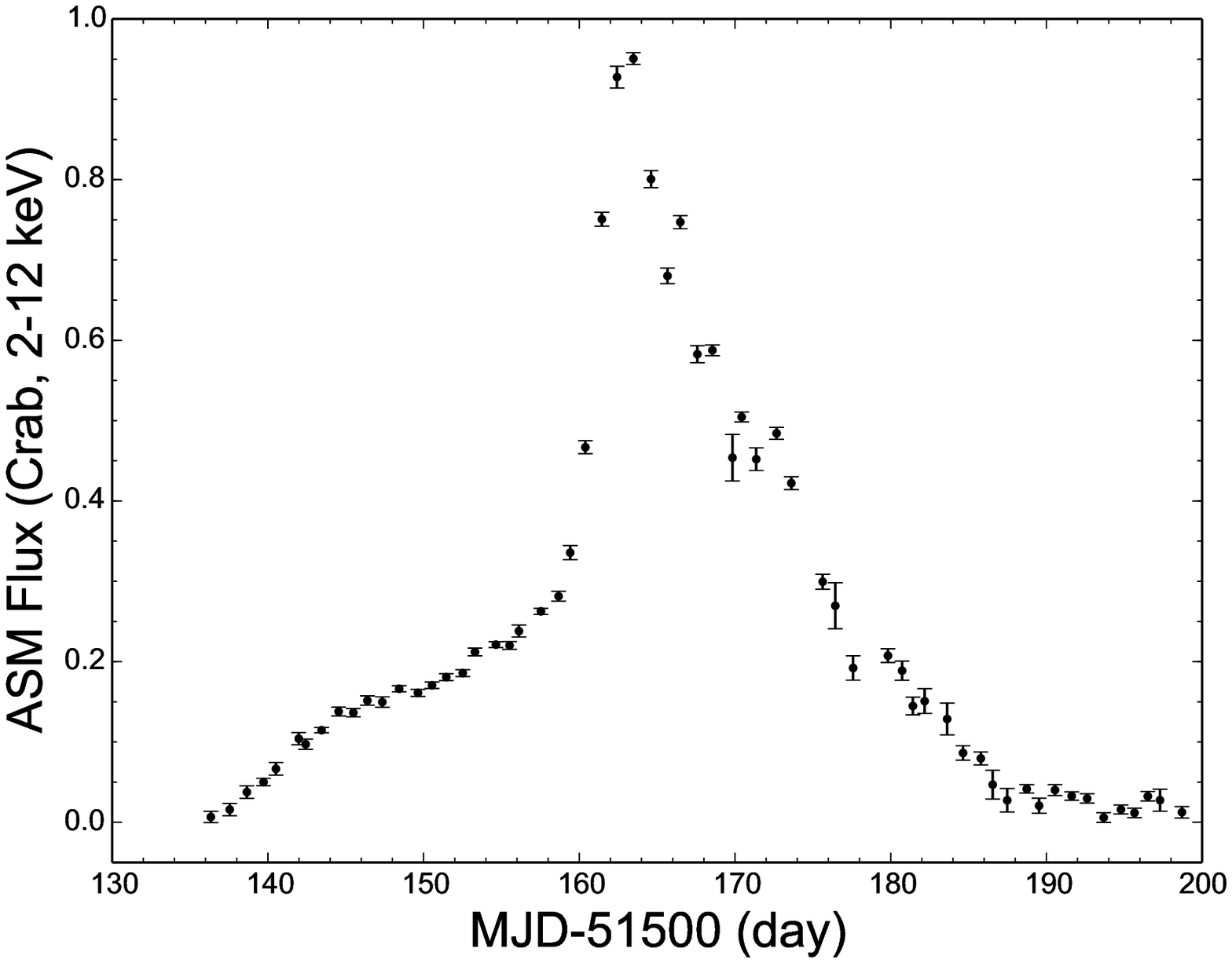}
\caption{\label{fg13}Light curves (with one day bin) of the black hole LMXB XTE J1550--564 during the 1998/1999 (Left panel) and 2000 (Right panel) outbursts in the range 2-12 keV as observed with {\it{RXTE}}/ASM (see section \ref{xte}).\label{fig13}}
\end{figure*}
The Galactic microquasar XTE J1550--564, which is one of the most studied Galactic X-ray transients, was discovered in 1998 with {\it{RXTE}}/ASM \citep{Smith,Sobczak}. This LMXB has shown five outbursts (1998/1999, 2000, 2001, 2001/2002, 2003) so far \citep{Tetarenko}, and some of them are `hard-only' outbursts (last three). For calculating the duty cycle, which comes roughly to be $0.04$, we only consider those outbursts in this paper for which the soft state has shown up in the X-ray spectra.
From modelling the optical and infrared observations of this black hole, \cite{Orsoz1} estimated the mass and distance of the black hole to be $9.1\pm 0.6\ M_{\odot}$ and $D=4.38\ \rm kpc$ respectively. Besides, the spin of the black hole was predicted to be $0.55^{+0.10}_{-0.15}$ by iron K$\alpha$ line method \citep{Steiner3}, and $0.34^{+0.20}_{-0.28}$ by using the continuum X-ray spectral fitting method \citep{Steiner3}. 

For the computation of alignment time of the black hole, we separately consider both the outbursts of 1998/1999 \citep{Smith,Sobczak} and 2000 \citep{Tomsick, Miller1} as they had sharply different accretion rate profiles (see the light curves of these outbursts in Fig. \ref{fg13}). During the 1998/1999 outburst, the X-ray flux rose to 6.48 Crab on MJD 51075, then fell nearly to its pre-flare flux, and rose again to 2.74 Crab on MJD 51201. Hence, it has undergone two successive outbursts during MJD 51062--51152 and MJD 51153--51297 (X-ray flux remained within $1\%$ of peak X-ray flux during both of these periods). In this case, we fit the two outburst accretion rate profiles with two separate triangular profiles, and consider both of the outbursts together as a single outburst event. But, the 2000 outburst (MJD 51637--51698) was comparatively much weaker, the X-ray flux rose to 0.95 Crab on MJD 51663, and taking this outburst as the prototype for all the outbursts gives a relatively longer alignment time. Hence, considering two distinctly different outbursts as prototypes, and combining the results from both, make our analysis more reliable. One can calculate the peak X-ray luminosity from the observed peak X-ray flux and the distance to the black hole, and it comes around $0.275\ L_{\rm Edd}$ for the first peak X-ray flare and $0.116\ L_{\rm Edd}$ for the second peak X-ray flare for the 1998/1999 outburst. Similarly, for the 2000 outburst, the peak X-ray luminosity comes around $0.0425\ L_{\rm Edd}$. The alignment time $t_{\rm a}$ comes roughly $0.3-1.1$ Gyr if one considers 2000 outburst profile as the prototype, whereas $t_{\rm a}$ comes around $0.034-0.082$ Gyr if one considers the outburst profile of 1998/1999 as the standard outburst profile for this black hole for the range of viscosity component $\nu_{20}$ mentioned in the section \ref{param} (corresponding range of $R_w$ is roughly $8-271$ $R_g$; see the equation (\ref{warprad}) for the definition of $R_w$). The age of this binary was estimated to be $4-13.5$ Gyr \citep{Fragos}. Therefore, the black hole is aligned with the outer disc for this source.

Although the microquasar XTE J1550--564 had been estimated to be misaligned by atleast $14^{\circ}$ \citep{Fragile}, \cite{Steiner} predicted that this black hole has most likely aligned its spin axis with the outer disc by modelling the X-ray and radio jet data for this black hole, and also gave an upper limit to the misalignment $<12^{\circ}$ up to $90\%$ confidence.  
Our analysis supports the prediction of \cite{Steiner}.  

\subsection{Alignment time: A typical Galactic LMXB}
We now estimate the alignment time for a typical Galactic transient LMXB with the average properties exhibited by them during outbursts in the {\it{RXTE}} era as reported by \cite{Yu}. We thus consider a black hole transient system, with the black hole mass of $9\ M_{\odot}$, undergoing an outburst of duration $88$ days with a peak X-ray luminosity of $0.052\ L_{\rm Edd}$ \citep{Yu} and duty cycle of $0.02$. The decay time of the accretion rate is three times greater than that of the rise time during the outburst. Using all these information, we find the alignment time takes the value $0.27-3.9\ \rm Gyr$ for the Kerr parameter range $0.1-0.9$ and the range of the viscosity component $\nu_{20}$ as mentioned in the section \ref{param} (corresponding range of $R_w$ is $4-720$ $R_g$; see the equation (\ref{warprad}) for the definition of $R_w$). This value is much longer than the previously estimated value of $10^6-10^8\ \rm yr$ \citep{Martin,Maccarone,Steiner}. Hence, one expects that some black hole LMXBs to be misaligned, and this can also affect the measurement of black hole spin using the continuum fitting method for the Galactic accreting black holes.

\subsection{Alignment time: Discussion}
Our formalism for the computation of alignment time (or time-scale) in transient accretion is valid if the viscous time-scale (in which matter is accreted through the disc) is less than the outburst rise time (as it is faster than the decay time). To check this, we estimate the viscous time scale for the part of the disc, which is primarily responsible for the black hole alignment process. We find that the main contribution of the component of LT torque responsible for the alignment process (i.e., $x$ component of the integral given in equation \ref{dj0}) comes from the region close to the warp radius (similar observation was mentioned earlier in \cite{Perego}; see their Fig.5). Away from the warp radius, the integral contribution falls rapidly, and becomes negligibly small far from this radius. Therefore, for calculating the alignment time for the Galactic black hole X-ray binaries mentioned in our paper, computing the integral only up to few times of $R_w$ is required. 
For this part of the accretion disc, which is primarily
responsible for the black hole alignment, a typical value
of the viscous time-scale for our parameter values is roughly 2 days. This is less than the observed outburst rise time (7-40 days) for the black holes we consider in our paper.

As we have seen earlier in the section \ref{per}, the contribution of the inner disc is captured by the quantity $R_{\rm in}/R_w$. For all the Galactic LMXBs considered in our study (including the case of a typical Galactic LMXB), the minimum value of $R_w$ considered for these sources is close to $R_{\rm in}$, and hence the inner disc may contribute significantly for atleast some ranges of $R_w$ starting from the minimum value.
 
In our work, we assume the viscosities to be radially constant. If one assumes the radial power law form of viscosities, the time-scales would have been higher than the constant viscosity case (roughly $1.5$ times the constant viscosity case for the Shakura-Sunyaev power law form) \citep{Martin}. Hence, even had we used a more realistic power law form of viscosities, as considered in the usual Shakura-Sunyaev model \citep{Shakura}, our conclusions regarding the misalignment of black hole spin for the above mentioned LMXBs and a typical Galactic accreting black hole would not change. 
\section{Summary}\label{con}
In this work, we consider a thin viscous accretion disc around a Kerr black hole in which the outer disc is tilted with respect to the spin direction of the black hole. In such a system, the inner disc may also be tilted with respect to the black hole spin.
A tilted thin inner accretion disc is particularly useful for probing the strong gravity region, as the LT precession can affect the observed X-ray timing and spectral features. But, the inner part of the accretion disc, was proposed to become aligned with the spin of the black hole by the well known Bardeen-Petterson effect. However, recently, BCB19 and \cite{CB17} have reported that the inner accretion disc could remain significantly tilted for a reasonable range of parameter values, like the Kerr parameter of the system. BCB19 considered the contribution of the inner disc in their calculations, and chose realistic boundary conditions (like truncating the disc at the innermost stable circular radius and considering the effect of viscous torques at the inner edge) for solving the warped disc equations formulated by Pringle (equation (\ref{warp})). 

As the black hole exerts a torque on the accretion disc due to the Lense-Thirring precession, and tries to align the inner disc with its own spin direction, the same torque is also applied on it due to the conservation of angular momentum. As a result of which, the black hole ultimately aligns itself with the outer disc. We analytically and numerically calculate the total LT torque acting on the black hole, and calculate the alignment and precession time-scales for both transient and persistent accretion. In our work, we take into account the contribution of the inner disc motivated by the theoretical investigation of BCB19 and observation of \cite{Ingram}, and explore in which region of parameter space our results deviate from what have been proposed earlier. Besides, for the transient accretion, we model the evolution of accretion rate during outburst by a triangular profile, and also allow both the viscosities as well as surface density to change during the outburst phase. We summarise our results below:
\begin{enumerate}
\item We find that the alignment time-scale for both persistent and transient accretion behaves  differently below and above the the critical Kerr parameter value $a_T$. Below $a_T$, the alignment time-scale increases as the Kerr parameter decreases contrary to what was proposed earlier. This critical value is found to depend on the viscosity component $\nu_2$ and the mass of the black hole. Above $a_T$, the alignment time-scale increases with the Kerr parameter following the same trend proposed in the past. For persistent accretion and above $a_T$, it roughly follows the $a^{1/2}$ behaviour as predicted by SF96.
\item The precession time-scale, although follows the same qualitative trend as a function of the Kerr parameter as found earlier both for the transient and persistent accretion, has smaller value than the earlier estimated result (i.e., as estimated by SF96) below a critical Kerr parameter (similar in behaviour to $a_T$) value for the persistent accretion, and beyond this value it follows roughly the $a^{1/2}$ form. 
\item The ratio between the alignment and precession time-scales in both the transient and persistent accretion is much different from unity for smaller values of the Kerr parameter contrary to the prediction of SF96, and it slowly approaches unity as the Kerr parameter value increases.  
\item The alignment and precession time-scales in the transient case in general have a longer value than the time-scales calculated for the persistent accretion.
\item We use our formalism of transient accretion for checking whether the black holes in the low mass X-ray binaries 4U 1543--47, H 1743--322, and XTE J1550--564 have got sufficient time to align its spin with the outer disc. We find that the black hole in the LMXB 4U 1543--47 could remain misaligned as independently predicted earlier by \cite{Morningstar}. We also show that the black hole in the LMXB XTE J1550--564 has most likely aligned itself with the outer disc, which is consistent with the prediction of \cite{Steiner} by an independent method. We also give an estimation of age of the LMXB H 1743--322 assuming the black hole to be misaligned with the outer disc (as noted by \cite{Ingram}). 
\item Using the mean outburst properties of transiently accreting Galactic black hole LMXBs in the {\it{RXTE}} era (as reported by \cite{Yu}), we also estimate the alignment time for such a typical transient black hole system using our formalism, and find that the alignment time (the time at which the mismatch in orientation between the black hole spin and the disc outer edge becomes $<1\%$ of the initial value) can be $(0.27-3.9)\times 10^9$ yrs, which is significantly higher than what has been predicted in the past. Hence, some Galactic LMXBs  
could remain misaligned with the outer disc. This may have an important implication in the context of spin measurement of black holes in LMXBs using the continuum X-ray spectral fitting method, in which an a priori assumption of alignment of black hole spin with the outer disc is usually made. 
\end{enumerate} 
\section*{Acknowledgements}
We thank the anonymous referee for constructive comments which helped improve the clarity of the paper.
C. C. gratefully acknowledges support from the National Natural Science Foundation of
China (NSFC), Grant No. 11750110410.
This work has made use of data provided by the {\it{RXTE}}/ASM teams at MIT and GSFC.






\bsp	
\label{lastpage}
\end{document}